\documentclass[11pt,a4paper]{article}

\usepackage[T1]{fontenc}
\usepackage[utf8]{inputenc}
\usepackage{amsmath,amssymb,amstext,mathtools}
\usepackage{graphicx}
\usepackage[margin=1in]{geometry}
\usepackage[nospace]{cite}
\usepackage[colorlinks,linkcolor=blue,citecolor=blue,urlcolor=blue]{hyperref}

\numberwithin{equation}{section}

\makeatletter
\newcommand{\linkref}[2][]{%
  \def\lr@arg{#1}%
  \ifx\lr@arg\@empty(\ref{#2})\else#1~(\ref{#2})\fi}
\makeatother

\title{The global diffusion limit for the space-dependent\\
variable-order time-fractional diffusion equation}

\author{%
Christopher N. Angstmann$^{1}$, Daniel S. Han$^{1,*}$, Bruce I. Henry$^{1}$,\\
Boris Z. Huang$^{1}$ and Zhuang Xu$^{2}$\\[6pt]
\small $^{1}$School of Mathematics and Statistics, University of New South Wales,\\
\small Sydney, New South Wales, Australia\\
\small $^{2}$University of Cambridge, Cambridge, UK\\[6pt]
\small $^{*}$Corresponding author: \texttt{daniel.han@unsw.edu.au}\\
}

\date{\today}

\begin{document}

\maketitle

\begin{abstract}
The diffusion equation and its time-fractional counterpart can be obtained via the diffusion limit of continuous time random walks with exponential and heavy-tailed waiting time distributions. The space-dependent variable-order time-fractional diffusion equation is a generalization of the time-fractional diffusion equation with a fractional exponent that varies over space, modelling systems with spatial heterogeneity. However, there has been limited work on defining a global diffusion limit and an underlying random walk for this macroscopic governing equation, which is needed to make meaningful interpretations of the parameters for applications. Here, we introduce continuous time and discrete time random walk models that limit to the variable-order fractional diffusion equation via a global diffusion limit and space- and time-continuum limits. From this, we show how the master equation of the discrete time random walk can be used to provide a numerical method for solving the variable-order fractional diffusion equation. The results in this work provide underlying random walks and an improved understanding of the diffusion limit for the variable-order fractional diffusion equation, which is critical for the development, calibration and validation of models for diffusion in spatially inhomogeneous media with traps and obstacles.
\end{abstract}


\bigskip

\section{Introduction}

The passage from a random walk (RW) to the macroscopic diffusion equation \cite{einstein1905motion}, and from a continuous time random walk (CTRW) \cite{montroll1965random} to the fractional diffusion equation \cite{hilfer1995fractional,metzler2000random,compte1996stochastic}, are celebrated results that have greatly improved theoretical understanding and modelling of diffusive transport in physical, biological, chemical, engineering and financial systems.
The RW models underpinning these macroscopic diffusion equations are spatially homogeneous.
In the standard RW, particles jump a fixed jump length or a Gaussian distributed jump length, at regular time intervals.
In the CTRW, particles wait a time, distributed by a waiting time distribution, before jumping a distance, distributed by the jump length distribution.
A time-fractional derivative can be introduced in the diffusion equation if the waiting time density has a power law tail and a space-fractional derivative can be introduced if the jump length density has a power law tail \cite{metzler2000random}.
The waiting times and jump lengths are random variables, but the probability density functions are the same at each spatial location, and thus the RW models are spatially homogeneous in this sense.
This spatial homogeneity is manifest as a constant fractional order derivative in the corresponding macroscopic time-fractional diffusion equation.

The passage from the master equation description of a RW to a partial differential equation (PDE) description involves a diffusion limit.
This can be formulated as a scaling limit, between the jump lengths and the time interval between jumps, or the rate of jumps, that enables the representation of the evolution equation for the probability density function of the RW as a well-defined PDE in continuous space $x$ and continuous time $t$.
This limit is a global diffusion limit if one scaling limit applies for all $x$.
If a scaling limit leading to a PDE is obtained in a small neighbourhood of $x$, rather than globally, then it is referred to as a local diffusion limit.
In the latter case, a PDE description on an extended domain would require a system of PDEs on local domains together with matching boundary conditions.

Fractional diffusion equations are experimentally characterized in many natural systems by measuring a mean square displacement growing nonlinearly with time: $\langle x^2(t)\rangle \sim t^{\alpha}$, where $\alpha$ is the fractional exponent.
Recent examples of spatial heterogeneity can be found across a range of scales, including diffusion in: biomolecular microfluidic devices \cite{pieprzyk2016spatially}, spatially varying porous media \cite{bruna2015diffusion}, composite media \cite{stickler2011continuous}, transport in ligand-receptor systems \cite{marbach2022mass,zheng2024hopping}, intracellular transport \cite{bressloff2013stochastic,han2020deciphering,fedotov2021variable,waigh2023heterogeneous}, electric transport in amorphous semiconductors \cite{ambegaokar1971hopping,egami2012spacetime}, soils \cite{henri2023control} and stellar clusters \cite{roupas2021gravitational}.
It is natural to ask if diffusion in spatially inhomogeneous systems, including possible trapping with power law waiting times, could be modelled using space-dependent variable-order fractional derivatives such that the fractional exponent, $\alpha(x)$, is a function of position.
Generalizations of fractional differential equations to variable-order have been considered theoretically \cite{samko1993integration,lorenzo2002variable,sun2019review,patnaik2020applications} but limited in their application since a microscopic mechanism that leads to a space- or time-dependent fractional exponent was missing.

The space-dependent variable-order time-fractional diffusion equation is \cite{chechkin2005fractional,fedotov2012subdiffusive,angstmann2013continuous,ricciuti2017semi,straka2018variable,fedotov2019asymptotic,roth2020inhomogeneous,fedotov2021variable}
\begin{equation}
        \frac{\partial p(x,t)}{\partial t} = \frac{\partial^2}{\partial x^2} \left[ \kappa_{\alpha(x)} {}_{0}{\mathcal{D}}_t^{1-\alpha(x)} p(x,t)\right],
        \label{eq:vofde}
\end{equation}
where the Riemann--Liouville fractional derivative is defined for $0<\alpha(x)\leq1$ as
\begin{equation}
        {}_{0}{\mathcal{D}}_t^{1-\alpha(x)} {p(x,t)} = \frac{1}{\Gamma\left(\alpha(x)\right)} { \frac{\partial}{\partial t}} \int_{0}^{t} \frac{ {p(x,t')}}{\left(t-t'\right)^{1-\alpha(x)}}dt'.
        \label{eq:RLderivative}
\end{equation}
In \linkref[equation]{eq:vofde}, $p(x,t)$ is the probability density function (PDF) of finding a RW at position $x$ and time $t$ and $\kappa_{\alpha(x)}$ is a space-dependent fractional diffusion coefficient with dimensions $\left[ \kappa_{\alpha(x)}\right] = L^2 T^{-\alpha(x)}$, where $L$ is length and $T$ is time.
The spatial heterogeneity of the underlying stochastic process in \linkref[equation]{eq:vofde} is encapsulated by the position dependent fractional exponent $\alpha(x)$.
The constant fractional exponent case of \linkref[equation]{eq:vofde} is well established for modelling anomalous subdiffusion in spatially homogeneous media with traps and obstacles where the mean square displacement is $\langle x^2\rangle \sim 2\kappa_{\alpha} t^{\alpha}$ \cite{metzler2000random}.
When $\alpha=1$, \linkref[equation]{eq:vofde} reverts to the standard diffusion equation.

Previous works have shown that the time-fractional diffusion \linkref[equation]{eq:vofde} for constant fractional exponent can be derived from a space-continuum limit and a diffusion limit of a CTRW \cite{hilfer1995fractional,compte1996stochastic,metzler2000random}.
Similarly, \linkref[equation]{eq:vofde} can be derived from the combined space and time-continuum limit and diffusion limit of a discrete time random walk (DTRW) \cite{angstmann2015discrete,nichols2018subdiffusive}.
The key to both derivations is that a global limit exists such that $\kappa_{\alpha}$ was the same limit everywhere in space.
It is natural to ask if a global limit can be obtained for $\kappa_{\alpha(x)}$ that may vary at different points in space.
The existence of a global limit is important for motivating the space-dependent variable-order time-fractional diffusion equation as a valid modelling equation that arises from an underlying stochastic process.
In fact, a stochastic simulation study on calibrating time in fractional diffusion equation models for transport in crowded environments \cite{ellery2016modeling} highlights the importance of such a valid diffusion limit.
Recent work has also shown a connection between subdiffusion and L\'evy flights through fractional advection-diffusion equations that are controlled by transport coefficients \cite{wang2020fractional}.

With this in mind, we first explore the continuum and diffusion limits for the fractional diffusion \linkref[equation]{eq:vofde} for a constant fractional exponent in \S\ref{sec:constantExponent} using a global limit, contextualizing its importance and the difficulty associated with making the fractional exponent space-dependent.
Then in \S\S\ref{sec:CTRWapproach} and \ref{sec:DTRWapproach}, we derive \linkref[equation]{eq:vofde} by taking the diffusion limits of CTRW and DTRW master equations, respectively.
In \S\ref{sec:DTRWapproach}, we also show that \linkref[equation]{eq:vofde} can be easily extended to spatial dimensions greater than one using the DTRW master equation.
Section~\ref{sec:numericalMethods} presents a numerical method for solving \linkref[equation]{eq:vofde} and compares this to Monte Carlo simulations of the underlying RW.

\section{Underlying random walk for the fractional diffusion equation with constant fractional exponent}
\label{sec:constantExponent}

In this section, we first summarize the steps, including the diffusion limits, leading from a CTRW and a DTRW to the time-fractional diffusion equation, with fractional exponent $\alpha$.
We then describe the problems in attempting to extend the diffusion limits in these derivations to incorporate a space-dependent fractional exponent $\alpha(x)$.

\subsection{Derivation from continuous time random walk}

The fractional diffusion equation is obtained from the master equation of a CTRW if the waiting time density has a power law tail, an example being the Mittag--Leffler waiting time density \cite{gorenflo2008continuous}
\begin{equation}
        \psi(t)=\frac{t^{\alpha-1}}{\tau^\alpha}
        E_{\alpha,\alpha}\left(-\left(\frac{t}{\tau}\right)^\alpha\right), \quad 0<\alpha\le 1,
        \label{eq:MLwaitingtimes}
\end{equation}
where $\tau$ is the characteristic waiting time and
\begin{equation}
        E_{\alpha,\beta}(z) = \sum_{k=0}^{\infty} \frac{z^k}{\Gamma(\alpha k + \beta)}
\end{equation}
is the two-parameter Mittag--Leffler function.
In the asymptotic limit as $t\rightarrow\infty$, this waiting time density has the power law tail \cite{gorenflo2008continuous}
\begin{equation}
        \psi(t) \sim \frac{\Gamma(\alpha+1)\sin(\alpha\pi)}{\pi}\tau^{\alpha} t^{-\alpha-1},
\end{equation}
promoting longer waiting times, which corresponds to longer trapping, as $\alpha$ approaches zero.
It is important to note that the characteristic waiting time, $\tau$, is not the expected waiting time as the latter diverges.
In constructing a CTRW that performs unbiased nearest-neighbour jumps on a lattice with spacings $\Delta x$ after waiting a randomly distributed time drawn from \linkref[equation]{eq:MLwaitingtimes}, the master equation can be written as \cite{montroll1965random,metzler2000random}
\begin{equation}
        \frac{\partial P(x,t)}{\partial t} = - i(x,t) + \frac{1}{2}i(x-\Delta x,t) + \frac{1}{2}i(x+\Delta x,t),
        \label{eq:CTRWmaster}
\end{equation}
where $P(x,t)$ is the probability that a RW particle is at position $x$ at time $t$ and
\begin{equation}
        i(x,t) = \tau^{-\alpha} {}_{0}{\mathcal{D}}_{t}^{1-\alpha} P(x,t)
        \label{eq:integral_escape_rate}
\end{equation}
can be identified as escape rates.
The derivation of \linkref[equation]{eq:CTRWmaster} and \linkref{eq:integral_escape_rate} can be found in appendix~\ref{appA}.
Taking the space-continuum limit of \linkref[equation]{eq:CTRWmaster} as $\Delta x \rightarrow 0$, one obtains
\begin{equation}
        \frac{\partial p(x,t)}{\partial t} = \left( \lim\limits_{\Delta x\rightarrow 0} \frac{ \Delta x^2}{2 \tau^{\alpha}}\right) \frac{\partial^2}{\partial x^2} {}_{0}{\mathcal{D}}_{t}^{1-\alpha} p(x,t).
        \label{eq:CTRWmaster2}
\end{equation}
Then, the diffusion limit for \linkref[equation]{eq:CTRWmaster2} results in the scaling relation $\Delta x^2 \sim \tau^{\alpha}$ such that
\begin{equation}
        \lim\limits_{\substack{\Delta x\rightarrow 0 \\ \tau \rightarrow 0}} \frac{ \Delta x^2}{2 \tau^{\alpha}} = \kappa_{\alpha}.
\end{equation}
In this limit, \linkref[equation]{eq:CTRWmaster2} becomes the fractional diffusion equation
\begin{equation}
        \frac{\partial p(x,t)}{\partial t} = \kappa_{\alpha} \frac{\partial^2}{\partial x^2} {}_{0}{\mathcal{D}}_{t}^{1-\alpha} p(x,t),
        \label{eq:fde}
\end{equation}
for constant fractional exponent, $\alpha$.

\subsection{Derivation from discrete time random walk}

In a similar vein, the fractional diffusion equation can be obtained from the master equation of a DTRW with discrete waiting time probability mass function (PMF) that has a power law tail, an example being the Sibuya distribution \cite{angstmann2015discrete,nichols2018subdiffusive}
\begin{equation}
        \psi(n) = \frac{\alpha}{n}\prod_{k=1}^{n-1} \left(1-\frac{\alpha}{k}\right), \quad n > 0.
\end{equation}
The Sibuya waiting time distribution, $\psi(n)$ is the probability that a RW waits exactly $n$ discrete time steps before jumping, with the probability of surviving the $k$th step being $1-\alpha/k$.
The Sibuya distribution has a power law tail in the asymptotic limit, $n\rightarrow \infty$, as the number of steps becomes large \cite{christoph2000scaled}
\begin{equation}
        \psi(n) \sim \frac{\Gamma(\alpha+1)\sin(\alpha\pi)}{\pi}n^{-\alpha-1}.
\end{equation}
The master equation for the unbiased nearest neighbour jump DTRW is \cite{angstmann2015discrete}
\begin{equation}
        \begin{split}
                P(x,n\Delta t) -P(x,(n-1)\Delta t) ={}&
                \sum_{m=0}^{n-1} \frac{K(n-m)}{2}
                \left[
                 P(x-\Delta x,m\Delta t) + P(x+\Delta x,m\Delta t)
                \right.\\
                &\left. \vphantom{\frac{A}{B}} - 2P(x,m\Delta t)
                \right],
        \end{split}
        \label{eq:DTRWmaster}
\end{equation}
with
\begin{equation}
        K(n) = (-1)^n \binom{1-\alpha}{n} +\delta_{n,1} - \delta_{n,0}.
\end{equation}
Here, $P(x,n)$ is the probability that a RW particle is at position $x$ after $n$ steps.
Taking the continuum limit in space and time of \linkref[equation]{eq:DTRWmaster} as $\Delta x \rightarrow 0$ and $\Delta t\rightarrow 0$, we obtain
\begin{equation}
        \frac{\partial p(x,t)}{\partial t} = \left(\lim\limits_{\substack{\Delta x\rightarrow 0 \\ \Delta t \rightarrow 0}} \frac{\Delta x^2}{2 \Delta t^{\alpha}} \right) \frac{\partial^2}{\partial x^2} {}_{0}{\mathcal{D}}_{t}^{1-\alpha} p(x,t).
        \label{eq:DTRWmasterContinuum}
\end{equation}
The diffusion limit for \linkref[equation]{eq:DTRWmasterContinuum} is
\begin{equation}
        \lim\limits_{\substack{\Delta x\rightarrow 0 \\ \Delta t \rightarrow 0}}\frac{\Delta x^2}{2 \Delta t^{\alpha}} = \kappa_{\alpha},
        \label{eq:DTRWmasterContinuumDiffCoeff}
\end{equation}
with the scaling relation $\Delta x^2 \sim \Delta t^{\alpha}$ and so \linkref[equation]{eq:DTRWmasterContinuum} is equivalent to \linkref[equation]{eq:fde}.
In the DTRW, the step time interval $\Delta t$ can be identified as a characteristic time and so the space and time-continuum limits are taken simultaneously with the diffusion limit.
The CTRW master \linkref[equation]{eq:CTRWmaster} has been extended to include forcing through biased jumps leading to fractional Fokker--Planck equations \cite{metzler2000random,henry2010fractional,angstmann2013continuous,angstmann2015generalized} and it has been extended to include reactions leading to fractional reaction diffusion equations \cite{henry2000fractional,sokolov2006reaction,henry2006anomalous,fedotov2010non,yuste2010reaction,angstmann2013continuous}. The DTRW has also been extended to include forcing \cite{angstmann2015discrete} and reactions \cite{angstmann2016stochastic}.

\subsection{Problems arising from a space-dependent fractional exponent}

Moving away from the assumption of spatial homogeneity represented via a constant fractional exponent $\alpha$, there has also been consideration of CTRWs for spatially inhomogeneous systems \cite{chechkin2005fractional,fedotov2012subdiffusive,angstmann2013continuous,ricciuti2017semi,straka2018variable,sun2019review,fedotov2019asymptotic,roth2020inhomogeneous,fedotov2021variable}.
In these previous works, a CTRW model with space-dependent waiting time densities characterized by the Mittag--Leffler density with a space-dependent fractional exponent $\alpha(x)$ was proposed and represented by a variable-order time-fractional diffusion equation with a space-dependent fractional-order time derivative \linkref{eq:vofde}.
In other words, \linkref[equation]{eq:fde} was transformed into \linkref[equation]{eq:vofde} through simply allowing $\alpha \rightarrow \alpha(x)$.
However, care must be taken when introducing spatial dependence into the fractional exponent as the global diffusion limit in the constant fractional exponent case must now be taken locally at all points in space \cite{roth2020inhomogeneous}.
This is evident from \linkref[equation]{eq:CTRWmaster2} and \linkref[equation]{eq:DTRWmasterContinuum} since the diffusion limits of both equations change when $\alpha\rightarrow\alpha(x)$,
\begin{equation}
        \begin{split}
                \text{(CTRW) } \lim\limits_{\substack{\Delta x\rightarrow 0 \\ \tau \rightarrow 0}} \frac{ \Delta x^2}{2 \tau^{\alpha}} &= \kappa_{\alpha} \rightarrow \lim\limits_{\substack{\Delta x\rightarrow 0 \\ \tau \rightarrow 0}} \frac{ \Delta x^2}{2 \tau^{\alpha(x)}} = \kappa_{\alpha(x)}\\
                \text{and (DTRW) }\lim\limits_{\substack{\Delta x\rightarrow 0 \\ \Delta t \rightarrow 0}}\frac{\Delta x^2}{2 \Delta t^{\alpha}} & = \kappa_{\alpha} \rightarrow \lim\limits_{\substack{\Delta x\rightarrow 0 \\ \Delta t \rightarrow 0}}\frac{\Delta x^2}{2 \Delta t^{\alpha(x)}} = \kappa_{\alpha(x)}.
        \end{split}
        \label{eq:localDiffusionLimits}
\end{equation}
The limit equations on the right-hand side of \linkref[equation]{eq:localDiffusionLimits} cannot hold globally if $\Delta x$, and $\tau$ or $\Delta t$, are constants with respect to $x$. To see this, suppose that $x^*\in \text{argmin}\left(\alpha(x)\right)$ and
\begin{equation}
        \kappa_{\alpha(x^*)} = \lim\limits_{\substack{\Delta x\rightarrow 0 \\ \tau \rightarrow 0}} \frac{\Delta x^2}{2\tau^{\alpha(x^*)}} \in \mathbb{R}.
\end{equation}
It then follows that if $x\notin \text{argmin}\left(\alpha(x)\right)$ and $\Delta x$ and $\tau$ are constants, then
\begin{equation}
        \kappa_{\alpha(x)} = \lim\limits_{\substack{\Delta x\rightarrow 0 \\ \tau \rightarrow 0}} \frac{\Delta x^2}{2\tau^{\alpha(x)}} = \lim\limits_{\tau \rightarrow 0} \frac{\kappa_{\alpha(x^*)}}{\tau^{\alpha(x)-\alpha(x^*)}} \rightarrow \infty.
        \label{eq:BadDiffCoeff}
\end{equation}
It is tempting to attempt achieving global limits in \linkref[equation]{eq:localDiffusionLimits} by simply replacing the quantities $\Delta x$, $\tau$ and $\Delta t$ by functions $\Delta x(x)$, $\tau(x)$ and $\Delta t(x)$, respectively.
This is naive because in \linkref[equation]{eq:localDiffusionLimits}, the quantity $\Delta x$ is a set length scale for space, effectively a `yardstick' and $\Delta t$ and $\tau$ are set time scales, effectively `clock time intervals'.
These physical quantities cannot be taken to depend on the location in space.
Note that even if $\kappa_{\alpha(x)}$ is constant in this formulation, it is still required to have space-dependent dimensions $L^2 T^{-\alpha(x)}$ to balance the spatial Laplacian and the space-dependent variable-order time-fractional derivative in \linkref[equation]{eq:vofde}.
This is physically problematic on a fundamental level and from a modelling perspective.

This physical dilemma was the primary motivation behind Straka's derivation of the variable-order fractional diffusion equation in \cite{straka2018variable} through embedding a CTRW process, a semi-Markov process, into a continuous time Markov chain before taking the diffusion limit.
In doing so, a global scaling parameter can be introduced to embed the Markov chain into continuous time.
While an elegant solution to the problem of local diffusion limits, this approach does not provide a clear physical understanding of this diffusion limit in terms of physically measurable quantities.
Roth and Sokolov considered the variable-order fractional diffusion equation with a locally defined diffusion coefficient \cite{roth2020inhomogeneous}
\begin{equation}
        \kappa_{\alpha(x)} = \lim\limits_{\substack{\Delta x\rightarrow 0 \\ \tau(x) \rightarrow 0}} \frac{\Delta x^2}{2\tau(x)^{\alpha(x)}}.
\end{equation}
Then by introducing a scaling parameter $\lambda$, they pointed out that the diffusion limit could be achieved locally by rescaling $\Delta x' = \lambda \Delta x$ and $\tau'(x) = \lambda^{2/\alpha(x)}\tau(x)$.
Roth and Sokolov also suggest that this equation could be calibrated and validated by short time measurements in local domains where $\kappa_{\alpha(x)}$ and $\alpha(x)$ may be considered as constant.

While this derivation may be mathematically valid, a variable-order fractional diffusion equation that is locally defined presents practical challenges to modelling, which is why a global approach is much needed.
In defining local diffusion limits, one needs to then define boundary conditions for each local region for what happens when the density of diffusing particles leaves one region and enters another \cite{roth2020inhomogeneous}.
While this serves as an approximation for slowly varying $\alpha(x)$, this approximation becomes intractable and unwieldy when $\alpha(x)$ varies sufficiently quickly.
Moreover, the local limits inherently define regions where $\alpha(x)$ is constant and so, for a continuously varying $\alpha(x)$ in space, an infinite number of these regions would be required to model the system.
An infinite number of boundary conditions to model the RWs in a finite domain would indeed be problematic!
Experimental examples of continuously varying $\alpha(x)$ have been shown in cellular systems for the lysosomal movement \cite{fedotov2021variable} and simulation studies of charged carrier transport in disordered semiconductors \cite{sibatov2009fractional,sibatov2024variable}.
Moreover, as we will show in \linkref[equations]{eq:badKalpha} and \linkref{eq:badKalpha2} and the surrounding discussion, the local limit for such an arbitrary function becomes theoretically problematic.
More recently, a rigorous treatment of the diffusion limits showed that a discrete space CTRW would generate space-dependent variable-order fractional derivatives in the diffusion limit \cite{kolokoltsov2023ctrw}.

\section{Continuous time random walk approach to a global diffusion limit}
\label{sec:CTRWapproach}

In this section, we first show how a diffusion equation with a space-dependent diffusion coefficient can be derived from a CTRW with nearest-neighbour jumps and a modified exponential waiting time density including a dimensional scale factor $\tau$ that can be applied globally.
We then show how a space-dependent variable order time-fractional diffusion equation can be obtained in the diffusion limit of a CTRW with an appropriately modified Mittag--Leffler waiting time density including a dimensionless scale factor $\tau$, that can be applied globally, and with a constant characteristic time scale $t_0$, with the dimensions of time.
We also provide some discussion around the problems in attempting to obtain well-defined diffusion limits with a space-dependent scale factor $\tau(x)$.
The result that will be derived in this section corresponds with Straka's result in \cite{straka2018variable} and leads to a globally defined diffusion coefficient.
We consider the master equation for a generalized CTRW on a lattice with $\Delta x$ spacings and unbiased nearest neighbour jumps (see for example \cite{angstmann2015generalized}),
\begin{equation}
        \begin{split}
                \frac{\partial P(x,t)}{\partial t}
                ={}
                &
                \frac{1}{2}\int_0^t K(x-\Delta x,t-t')P(x-\Delta x,t') dt'
                +\frac{1}{2}\int_0^t K(x+\Delta x,t-t')P(x+\Delta x,t')\, dt'
                \\
                &
                -\int_0^t K(x,t-t')P(x,t')\, dt',
        \end{split}
        \label{eq:ctrwMaster}
\end{equation}
where the kernel is defined in Laplace space by
\begin{equation}
        \mathcal{L}_t \left\{ K(x,t) \right\}(s) = \frac{       \mathcal{L}_t \left\{ \psi(x,t) \right\} (s)}{        \mathcal{L}_t \left\{ \Psi(x,t) \right\} (s)}.
        \label{eq:LaplaceSpaceKernel}
\end{equation}
In \linkref[equation]{eq:LaplaceSpaceKernel}, $\psi(x,t)$ is a space-dependent waiting time density, $\Psi(x,t) = 1-\int_{0}^{t}\psi(x,t')dt'$ is a space-dependent survival probability and $\mathcal{L}_t\left\{ \cdot \right\}(s)$ is the Laplace transform with respect to time $t$ and transform variable $s$.
Note that \linkref[equation]{eq:ctrwMaster} is equivalent to \linkref[equation]{eq:CTRWmaster} when the waiting time density is the Mittag--Leffler waiting time density; see \linkref[equation]{eq:MLwaitingtimes}.

\subsection{Space-dependent exponential waiting time density}

To demonstrate a well-known global diffusion limit, we consider the Markovian case of \linkref[equation]{eq:ctrwMaster} with space-dependent exponential waiting time density
\begin{equation}
        \psi(x,t) = \frac{\lambda(x)}{\tau} e^{-\frac{\lambda(x)}{\tau}t}, \quad \text{for}\quad 0<\lambda(x) \leq 1.
        \label{eq:RescaledExponential}
\end{equation}
Then, the kernel in \linkref[equation]{eq:ctrwMaster} defined by \linkref[equation]{eq:LaplaceSpaceKernel} is
\begin{equation}
        K(x,t) = \frac{\lambda(x)}{\tau}\delta (t),
        \label{eq:exponentialKernel}
\end{equation}
where $\delta(t)$ is the Dirac delta function.
Substituting \linkref[equation]{eq:exponentialKernel} into the master \linkref[equation]{eq:ctrwMaster}, we obtain
\begin{equation}
                \frac{\partial P(x,t)}{\partial t}
                =\frac{\lambda(x-\Delta x)}{2\tau}P(x-\Delta x,t)
                +\frac{\lambda(x+\Delta x)}{2\tau}P(x+\Delta x,t) -\frac{\lambda(x)}{\tau}P(x,t).
        \label{master_markov}
\end{equation}
In the diffusion limit as $\Delta x \rightarrow 0$ and $\tau\rightarrow 0$, \linkref[equation]{master_markov} becomes
\begin{equation}
        \frac{\partial p(x,t)}{\partial t} = \frac{\partial^2}{\partial x^2} \left[D(x)p(x,t)\right],
        \label{eq:CTRWMarkov_continuummaster}
\end{equation}
where the space-dependent diffusion coefficient is
\begin{equation}
        D(x) = \lim_{\substack{\Delta x\to 0\\\tau\to 0}}\frac{\Delta x^2}{2\tau} \lambda(x).
        \label{eq:MarkovianSpaceDependentDiffusionCoeff}
\end{equation}
Although there are different waiting time densities at each position, the diffusion limit can be taken globally since $\lambda(x)$ does not depend on the parameters taken in the limit, $\Delta x$ and $\tau$.
Note that equivalently, we could have introduced $\tau(x) = \tau/\lambda(x)$ and obtained \linkref[equation]{eq:MarkovianSpaceDependentDiffusionCoeff}.

\subsection{Space-dependent Mittag--Leffler waiting time density}

Now, we move to the non-Markovian case when the waiting time density is defined as the space-dependent Mittag--Leffler waiting time density similar to \linkref[equation]{eq:MLwaitingtimes},
\begin{equation}
        \psi(x,t)=\frac{t^{\alpha(x)-1}}{\tau^{\alpha(x)}}
        E_{\alpha(x),\alpha(x)}\left(-\left(\frac{t}{\tau}\right)^{\alpha(x)}\right),
        \label{eq:MLwaitingtimes_spacedep}
\end{equation}
with $0<\alpha(x) \leq 1$.
This approach has been taken in previous works and below we show the problematic diffusion limit using this approach \cite{chechkin2005fractional,korabel2010paradoxes,fedotov2012subdiffusive,fedotov2019asymptotic}.
Using \linkref[equation]{eq:MLwaitingtimes_spacedep}, \linkref[equation]{eq:ctrwMaster} becomes
\begin{equation}
        \begin{split}
                \frac{\partial P(x,t)}{\partial t}
                ={}
                &
                \frac{1}{2\tau^{\alpha(x-\Delta x)}} {}_{0}{\mathcal{D}}_{t}^{1-\alpha(x-\Delta x)} P(x-\Delta x,t)
                +\frac{1}{2\tau^{\alpha(x+\Delta x)}} {}_{0}{\mathcal{D}}_{t}^{1-\alpha(x+\Delta x)} P(x+\Delta x,t)
                \\
                &
                -\frac{1}{\tau^{\alpha(x)}} {}_{0}{\mathcal{D}}_{t}^{1-\alpha(x)}P(x,t),
        \end{split}
        \label{eq:master_nonmarkov}
\end{equation}
where ${}_{0}{\mathcal{D}}_{t}^{1-\alpha(x)}$ is the Riemann--Liouville fractional derivative as defined in \linkref[equation]{eq:RLderivative}.
In the diffusion limit as $\Delta x \rightarrow 0$ and $\tau\rightarrow0$, \linkref[equation]{eq:master_nonmarkov} is assumed to become the variable-order fractional diffusion equation; see \linkref[equation]{eq:vofde}.
However, the limit from \linkref[equation]{eq:master_nonmarkov} to \linkref[equation]{eq:vofde} implies that the fractional diffusion coefficient is defined as
\begin{equation}
        \kappa_{\alpha(x)} = \lim_{\substack{\Delta x\to 0\\\tau\to 0}} \frac{\Delta x^2}{2\tau^{\alpha(x)}}.
        \label{eq:WrongFractionalDiffCoeff}
\end{equation}
As previously shown by \linkref[equation]{eq:BadDiffCoeff}, the limit in \linkref[equation]{eq:WrongFractionalDiffCoeff} is unphysical as the diffusion limit must be taken differently dependent on position, diverging for positions where the fractional exponent is not a minimum.
This motivates using the rescaled space-dependent Mittag--Leffler waiting time density,
\begin{equation}
        \psi(x,t)=\frac{t^{\alpha(x)-1}}{\tau t_0^{\alpha(x)}}
        E_{\alpha(x),\alpha(x)}\left(-\frac{1}{\tau}\left(\frac{t}{t_0}\right)^{\alpha(x)}\right),
        \label{eq:MLwaitingtimes_rescaledspacedep}
\end{equation}
where $t_0$ is a characteristic time and $\tau$ is a dimensionless scale factor.
This is analogous to rescaling the intensity in \linkref[equation]{eq:RescaledExponential} by $\tau$ and $\lambda(x)$.
Using \linkref[equation]{eq:MLwaitingtimes_rescaledspacedep}, the master \linkref[equation]{eq:ctrwMaster} can be written, in a similar way to \linkref[equation]{eq:master_nonmarkov},
\begin{equation}
        \begin{split}
                \frac{\partial P(x,t)}{\partial t} ={}
                &
                \frac{1}{2\tau t_0^{\alpha(x-\Delta x)}} {}_{0}{\mathcal{D}}_{t}^{1-\alpha(x-\Delta x)} P(x-\Delta x,t)
                +\frac{1}{2\tau t_0^{\alpha(x+\Delta x)}} {}_{0}{\mathcal{D}}_{t}^{1-\alpha(x+\Delta x)} P(x+\Delta x,t)
                \\
                &
                -\frac{1}{\tau t_0^{\alpha(x)}} {}_{0}{\mathcal{D}}_{t}^{1-\alpha(x)}P(x,t).
        \end{split}
        \label{eq:master_rescalednonmarkov}
\end{equation}
Clearly, the only difference between \linkref[equations]{eq:master_nonmarkov} and \linkref{eq:master_rescalednonmarkov} is the introduction of an additional parameter $t_0$.
In \linkref[equation]{eq:MLwaitingtimes_spacedep}, $\tau$ is the characteristic time but, in \linkref[equations]{eq:MLwaitingtimes_rescaledspacedep}, $t_0$ is the characteristic time and $\tau$ is now a dimensionless scale factor.
In the diffusion limit, \linkref[equation]{eq:master_rescalednonmarkov} becomes \linkref[equation]{eq:vofde} with the diffusion coefficient defined as
\begin{equation}
        \kappa_{\alpha(x)} = \lim_{\substack{\Delta x\to 0\\\tau\to 0}} \frac{\Delta x^2}{2\tau }t_0^{-\alpha(x)} = Dt_0^{-\alpha(x)}.
        \label{eq:RightFractionalDiffCoeff}
\end{equation}
It is clear from comparing \linkref[equations]{eq:WrongFractionalDiffCoeff} and \linkref{eq:RightFractionalDiffCoeff} that the rescaled waiting time density \linkref{eq:MLwaitingtimes_rescaledspacedep} generates a global diffusion limit for the fractional diffusion equation that is analogous to \linkref[equations]{eq:MarkovianSpaceDependentDiffusionCoeff}.
The rescaling in \linkref[equation]{eq:RightFractionalDiffCoeff} is equivalent to that obtained by Straka in \cite{straka2018variable} but here we have obtained the result directly from a CTRW by using a rescaled waiting time density. Straka obtained the scaling result by embedding a CTRW into a continuous time Markov chain.
While a global diffusion limit is now well defined in the CTRW framework, doing so has introduced a new dimensionless time-related parameter, $\tau$, the underlying question of what this parameter physically represents still remains.
To answer this, we derive the space-dependent variable-order time-fractional diffusion equation from a DTRW framework.

Before leaving this section, we note that it may be tempting to consider a more general waiting time density
\[
\psi(x,t)= \frac{t^{\alpha(x)-1}}{\tau(x)^{\alpha(x)}} E_{\alpha(x),\alpha(x)}
\left(
-\left(
\frac{t}{\tau(x)}
\right)^{\alpha(x)}
\right)
\]
in the master equation and then take a continuum limit and a diffusion limit with a diffusion coefficient defined through
\begin{equation}
        K_{\alpha}(x)=\lim\limits_{\substack{\Delta x\rightarrow 0 \\ \tau(x) \rightarrow 0}}
        \frac{\Delta x^2}{2\tau(x)^{\alpha(x)}}.
        \label{eq:badKalpha}
\end{equation}
However, it is not appropriate to take a limit $\tau(x)\to 0$ for an arbitrary function $\tau(x)$ over all space in this limit.
Such a limit could be defined locally at a particular value $x=x_0$ where
\begin{equation}
        K_\alpha(x_0)=
        \lim\limits_{\substack{\Delta x\rightarrow 0 \\ \tau(x_0) \rightarrow 0}}
        \frac{\Delta x^2}{\tau(x_0)^{\alpha(x_0)}}
        \label{eq:badKalpha2}
\end{equation}
but in this case, the diffusion coefficient and anomalous exponent are constant values locally, as considered in \cite{roth2020inhomogeneous}.

The limit $\tau(x)\to 0$ in \linkref[equation]{eq:badKalpha} can be taken in a variety of ways. The simplest way is through the introduction of a scaling factor $\tau_0$ that does not depend on $x$, with $\tau(x)=\tau_0 T(x)$.
The limit $\tau(x)\to 0$ can then be taken by taking the limit $\tau_0\to 0,$ but then it is evident that the same problem surrounding \linkref[equation]{eq:BadDiffCoeff} appears.
If we want some suitable function of $\tau_0$, $\tau(x) = f(\tau_0; x)T(x)$, instead of the simplest scaling, then this case will revert back to \linkref[equations]{eq:RightFractionalDiffCoeff} or to a more general case considered below in \linkref[equation]{eq:RightFractionalDiffCoeff_moregeneral} and \linkref{eq:RightFractionalDiffCoeff_evenmoregeneral}.
For any general function of $\tau_0$, it is clear from dimensional arguments that $T(x)$ must have units of time and $\tau_0$ becomes equivalent to the dimensionless scaling constant in our discussion surrounding \linkref[equation]{eq:RightFractionalDiffCoeff}.
It may be mathematically valid and tempting to rearrange \linkref[equation]{eq:badKalpha} and simply define $\tau(x)$ in terms of $\Delta x$ and $K_{\alpha}(x)$, but this would be describing a microscopic RW time scale in terms of a macroscopic diffusion coefficient.

These general considerations can be formulated with the rescaled space-dependent Mittag--Leffler waiting time density defined in \linkref[equation]{eq:MLwaitingtimes_rescaledspacedep} providing: a dimensionless scale factor $\tau$ that can be applied globally; a constant characteristic time scale $t_0$ with appropriate dimensions; and a well-defined limit with a differentiable diffusion coefficient defined in \linkref[equation]{eq:RightFractionalDiffCoeff}.
One could also consider the more general form
\begin{equation}
        \kappa_{\alpha(x)} = Dt_0^{-\alpha(x)} \Lambda\left(\frac{x}{x_0}\right)^{-\alpha(x)},
        \label{eq:RightFractionalDiffCoeff_moregeneral}
\end{equation}
where $\Lambda(x/x_0)$ is a general function of a rescaled unitless spatial variable $x/x_0$. However, for transparency on the RW origin of the $t_0$ factor, we consider the case $\Lambda(x/x_0)=1$ throughout this manuscript.
In fact, we can generalize even further and obtain something that is similar in form to the diffusion coefficent presented in \cite{roth2020inhomogeneous} while retaining physical meaning of the parameters.
We find that the most general form of the diffusion coefficient is
\begin{equation}
        \begin{split}
                \kappa_{\alpha(x)} &= \lim_{\substack{\Delta x\to 0\\\tau_0\to 0}} \frac{\Delta x^2}{2\tau(x) } t_0^{-\alpha(x)} \Lambda\left(\frac{x}{x_0}\right)^{-\alpha(x)}\\
                &=D(x)t_0^{-\alpha(x)} \Lambda\left(\frac{x}{x_0}\right)^{-\alpha(x)},
        \end{split}
        \label{eq:RightFractionalDiffCoeff_evenmoregeneral}
\end{equation}
where $\tau(x) = \tau_0 T(x)$.
Here, $D(x)$ represents the component of the diffusion coefficient that arises from heterogeneity in spatial scaling, i.e. the ratio $\Delta x^2/\left(\tau_0T(x)\right)$.
The overall diffusion coefficient is then obtained by multiplying the heterogeneous scaling, i.e. $t_0^{-\alpha(x)} \Lambda\left(x/x_0\right)^{-\alpha(x)}$.
This two component nature of the diffusion coefficient has recently been shown to be experimentally valid through fractionation patterns in spatially heterogeneous environments, albeit the experiments were observing purely diffusive tracers \cite{kim2024fractionation}.

\section{Discrete time random walk approach to a global diffusion limit}
\label{sec:DTRWapproach}

In this section, we first derive the master equation for a DTRW with a space-dependent waiting time PMF and an arbitrary transition PMF.
It is not possible to obtain a well-defined global diffusion limit if the transition PMF is limited to nearest-neighbour jumps.
Guided by our experience with the diffusion limit of the CTRW leading to the space-dependent variable-order time-fractional diffusion equation, we consider a modified transition PMF that introduces an additional parameter through the probability of either doing a self-jump or doing a nearest-neighbour jump after each waiting time.
We then show, in \S\ref{sec:DTRWmarkov}, how a diffusion equation with a space-dependent diffusion coefficient can be derived from this DTRW with a Markovian waiting time PMF.
In \S\ref{sec:DTRWnonmarkov}, we show how a space-dependent variable-order time-fractional diffusion equation can be obtained in the diffusion limit of this DTRW with a heavy-tailed non-Markovian PMF.

\subsection{Derivation of the discrete time random walk master equation}

We begin by deriving the DTRW framework \cite{angstmann2015discrete,angstmann2016stochastic} with a spatially inhomogeneous environment at mesoscopic scales. A single particle performing jumps at discrete times, $n\in\mathbb{N}$, on an infinite one-dimensional lattice $V=\{\cdots,i-1,i,i+1,\cdots\}$ with $i \in \mathbb{Z}$.
At each discrete time step, the particle either remains at the same location or transitions to a different location.
We define $\rho(j,n|i,m)$ as the transition PMF for a particle that arrived at lattice site $i$ at time step $m$ to transition to lattice site $j$ at time step $n$ greater than $m$.
In what follows, we assume that the jumps between sites and time steps are independent and so can be decoupled such that
\begin{equation}
        \rho(j,n|i,m) = w(j|i) \psi(i,n-m),
\end{equation}
where $w(j|i)$ is the PMF for transitioning to site $j$ given the particle was at site $i$ previously and $\psi(i,n-m)$ is the PMF for waiting $n-m$ time steps at site $i$ before transitioning.
Note that each PMF is normalized such that
\begin{equation}
        \sum_{j=-\infty}^{\infty}\sum_{n=0}^{\infty} \rho(j,n|i,m) = 1,
\end{equation}
with
\begin{equation}
        \sum_{n=0}^{\infty} \psi(i,n) = 1
        \quad \text{and} \quad
        \sum_{j=-\infty}^{\infty} w(j|i) = 1.
\end{equation}
In addition, we require that $\psi(i,0)=0$ meaning that no transitions can happen in less than one discrete time step.

The probability flux of a particle arriving at site $i$ at time step $m$ given an initial site $i_0$ at time step $n_0$ can be written as
\begin{equation}
        \begin{split}
                Q(i,m|i_0,n_0)
                &
                =
                \delta_{i,i_0}\delta_{m,n_0}
                +\sum_{j=-\infty}^{\infty}\sum_{k=0}^{m-1} \rho(i,m|j,k)Q(j,k|i_0,n_0),
                \\
                &
                =
                \delta_{i,i_0}\delta_{m,n_0}
                +\sum_{j=-\infty}^{\infty} \sum_{k=0}^{m-1} w(i|j)\psi(j,m-k)Q(j,k|i_0,n_0),
        \end{split}
        \label{eq:DTRWprobabilityFlux}
\end{equation}
with $Q(i,m|i_0,n_0) = 0$ for $m<n_0$.
The survival probability of a particle remaining at site $i$ at time step $n$ given that it arrived there at an earlier time step $m$ is
\begin{equation}
        \Psi(i,n-m) = 1- \sum_{k=0}^{n-m}\psi(i,k).
        \label{eq:DTRWsurvivalprobability}
\end{equation}
In combining the probability flux \linkref{eq:DTRWprobabilityFlux} and the survival probability \linkref{eq:DTRWsurvivalprobability}, the probability of finding the particle at site $i$ at time step $n$ can be written as
\begin{equation}
        P(i,n|i_0,n_0) = \sum_{m=0}^{n}\Psi(i,n-m)Q(i,m|i_0,n_0).
        \label{eq:DTRWChapmanKolmogorov}
\end{equation}
To derive the general master equation, the change in probability mass between two successive time steps $n-1$ and $n \geq 1$ is
\begin{align}
       & P(i,n|i_0,n_0)
        - P(i,n-1|i_0,n_0) = Q(i,n|i_0,n_0)
       \nonumber \\
        &\quad + \sum_{m=0}^{n-1} \left[\Psi(i,n-m)-\Psi(i,n-1-m)\right] Q(i,m|i_0,n_0).
        \label{eq:DTRWmaster1}
\end{align}
Then using \linkref[equation]{eq:DTRWsurvivalprobability}, the difference in survival probabilities can be expressed as
\begin{equation}
        \Psi(i,n-m)-\Psi(i,n-1-m) = -\psi(i,n-m).
        \label{eq:DTRWsurvival_pmf_relation}
\end{equation}
Substituting \linkref[equations]{eq:DTRWsurvival_pmf_relation} and \linkref{eq:DTRWprobabilityFlux} into \linkref[equation]{eq:DTRWmaster1}, we obtain
\begin{align}
        P(i,n|i_0,n_0) - P(i,n-1|i_0,n_0) ={}&
        \delta_{i,i_0}\delta_{n,n_0}
        +\sum_{j=-\infty}^{\infty}\sum_{m=0}^{n-1}w(i|j)\psi(j,n-m)Q(j,m|i_0,n_0)
        \nonumber \\
       & -\sum_{m=0}^{n-1}\psi(i,n-m)Q(i,m|i_0,n_0).
        \label{eq:DTRW_premaster}
\end{align}
In order to close \linkref[equation]{eq:DTRW_premaster} and obtain a DTRW master equation, we require an expression that gives $Q(i,n|i_0,n_0)$ in terms of $P(i,n|i_0,n_0)$.
To do this, we use the $\mathcal{Z}$-transform defined by
\begin{equation}
        \hat{Y}(i,z) = \mathcal{Z}_n\left\{ Y(i,n) \right\}(z) = \sum_{n=0}^{\infty}z^{-n}Y(i,n).
        \label{eq:Ztransform}
\end{equation}
Applying the $\mathcal{Z}$-transform to \linkref[equation]{eq:DTRWChapmanKolmogorov}, we obtain
\begin{equation}
        \hat{P}(i,z|i_0,n_0) = \hat{\Psi}(i,z) \hat{Q}(i,z|i_0,n_0).
        \label{eq:DTRWChapmanKolmogorovZtransformed}
\end{equation}
Further, the $\mathcal{Z}$-transform of the summation terms in \linkref[equation]{eq:DTRW_premaster} is
\begin{equation}
        \begin{split}
                \mathcal{Z}_n \left\{ \sum_{m=0}^{n-1} \psi(i,n-m) Q(i,m|i_0,n_0) \right\} (z) =
                \hat{\psi}(i,z) \hat{Q}(i,z|i_0,n_0).
        \end{split}
        \label{eq:DTRWfluxsumZtransformed}
\end{equation}
Combining \linkref[equations]{eq:DTRWChapmanKolmogorovZtransformed} and \linkref{eq:DTRWfluxsumZtransformed}, the sums in \linkref[equation]{eq:DTRW_premaster} can be expressed as
\begin{equation}
        \begin{split}
                \sum_{m=0}^{n-1} \psi(i,n-m)
                Q(i,m|i_0,n_0) =
                \sum_{m=0}^{n-1} K(i,n-m) P(i,m|i_0,n_0),
        \end{split}
        \label{eq:DTRWKernel}
\end{equation}
with the kernel $K(i,n)$ defined via the $\mathcal{Z}$-transform as
\begin{equation}
        \hat{K}(i,z) = \frac{\hat{\psi}(i,z)}{\hat{\Psi}(i,z)}.
        \label{eq:DTRWKernelZtransformed}
\end{equation}
Finally, the DTRW master equation is
\begin{equation}
        \begin{split}
                P(i,n)
                - P(i,n-1) =
                \sum_{j=-\infty}^{\infty}\sum_{m=0}^{n-1}w(i|j)K(j,n-m)P(j,m)
                -\sum_{m=0}^{n-1}K(i,n-m)P(i,m),
        \end{split}
        \label{eq:DTRW_generalmaster}
\end{equation}
where the above equation omits the initial conditions for convenience since we assume $n_0=0$ and $n\geq1$ without loss of generality.
For the unbiased nearest-neighbour jumps, the PMF for transitions between sites is
\begin{equation}
        w(i|j) = \frac{1}{2}\delta_{j,i+1} + \frac{1}{2}\delta_{j,i-1},
        \label{eq:DTRWwrongtransitionPMF}
\end{equation}
which makes \linkref[equation]{eq:DTRW_generalmaster} equivalent to \linkref[equation]{eq:DTRWmaster}.
Observing that the continuum and diffusion limits of \linkref[equation]{eq:DTRWmaster} lead to the fractional diffusion \linkref[equation]{eq:DTRWmasterContinuum} and the fractional diffusion coefficient \linkref{eq:DTRWmasterContinuumDiffCoeff}, it is clear that choosing the PMF as \linkref[equation]{eq:DTRWwrongtransitionPMF} will lead to the same unphysical, space-dependent, local diffusion limits analogous to \linkref[equation]{eq:WrongFractionalDiffCoeff}.

The natural question then arises: what is the form of $K(i,n)$ and $w(i|j)$ that generates the analogue of the CTRW diffusion coefficient with global scaling in \linkref[equation]{eq:RightFractionalDiffCoeff}?
In the continuous time framework, one could introduce an arbitrary parameter $\tau$ into the waiting time density \linkref{eq:MLwaitingtimes_rescaledspacedep} that influences time scaling such that a global diffusion limit can be defined.
Here, we are attempting to obtain a DTRW that limits to the variable-order time-fractional diffusion \linkref{eq:vofde} without the introduction of additional parameters in the continuum approximation, beyond those in the DTRW.
This leaves no other choice except to modify the site transition PMF \linkref{eq:DTRWwrongtransitionPMF}.
Since the global diffusion limit in the continuous case required introducing an additional time scaling parameter, we look to introducing an additional outcome, at each time step, of remaining at the same site in a jump such that the transition PMF becomes
\begin{equation}
        w(i|j) = \frac{1-r(j)}{2}\delta_{j,i+1} + \frac{1-r(j)}{2}\delta_{j,i-1} + r(j)\delta_{j,i}.
        \label{eq:DTRWrighttransitionPMF}
\end{equation}
Using \linkref[equation]{eq:DTRWrighttransitionPMF} in \linkref[equation]{eq:DTRW_generalmaster}, we obtain the general self-jumping DTRW master equation
\begin{equation}
        \begin{split}
                P(i,n)-P(i,n-1)=
                \sum_{m=0}^{n-1}\left[ \vphantom{\frac{A}{2} } \right.
                & \left. \frac{1-r(i-1)}{2}K(i-1,n-m)P(i-1,m)
                \right.
                \\
                &
                + \frac{1-r(i+1)}{2}K(i+1,n-m)P(i+1,m)
                                \\
                & \left. \vphantom{\frac{A}{2} }
                - \left(1-r(i)\right)K(i,n-m)P(i,m) \right].
        \end{split}
        \label{eq:DTRWmasterRestNN}
\end{equation}
Physically, the quantity $r(j)$ represents the probability of failure to leave site $j$ in the DTRW.

\subsection{Space-dependent Markovian waiting step mass function}
\label{sec:DTRWmarkov}

As before, we first consider the Markovian DTRW when the probability to jump from site $i$ on any time step is
\begin{equation}
        \psi(i,n) = \begin{cases}
                0, & \text{if } n\leq 0,\\
                \omega(i)(1-\omega(i))^{n-1}, & n>0.
        \end{cases}
        \label{eq:DTRWMarkovwaitingtime}
\end{equation}
\linkref[Equation]{eq:DTRWMarkovwaitingtime} gives the survival probability, a geometric distribution,
\begin{equation}
        \Psi(i,n) = (1-\omega(i))^n,
        \label{eq:DTRWMarkovsurvival}
\end{equation}
and kernel,
\begin{equation}
        K(i,n)=\omega(i)\delta_{n,1}.
        \label{eq:DTRWMarkovkernel}
\end{equation}
Using \linkref[equation]{eq:DTRWMarkovkernel} in \linkref[equation]{eq:DTRWmasterRestNN}, we obtain
\begin{equation}
        \begin{split}
                P(i,n)-P(i,n-1) ={}&\frac{1-r(i-1)}{2} \omega(i-1)P(i-1,n-1)
                +\frac{1-r(i+1)}{2} \omega(i+1)P(i+1,n-1)
                \\
                &
                -\big(1-r(i)\big)\omega(i)P(i,n-1).
        \end{split}
        \label{eq:DTRWMarkovianmaster}
\end{equation}

To consider the continuum limit, and diffusion limit, of the master equations, we associate the discrete function $Y(i,n)$ with a sequence of continuous functions, $y_\Delta(x,t)$, through the identification
\begin{equation}
        y_\Delta(i\Delta x,n\Delta t)=Y(i,n),
        \label{eq:seqcontfunc}
\end{equation}
for different possible lattice spacings, $\Delta x$, and time step intervals, $\Delta t$.
The continuum limit then identifies
\begin{equation}
        y(x,t)=\lim_{\substack{\Delta x\to 0\\\Delta t\to 0}}y_\Delta(i\Delta x,n\Delta t).
        \label{climit}
\end{equation}
It is straightforward to show that if the lattice spacing $\Delta x$ and the clock time interval $\Delta t$ are uniform parameters then in the continuum limit, $\Delta x\to 0, \Delta t\to 0$, the master equation for the DTRW with self-jumping can be reduced to
\begin{equation}
        \frac{\partial p(x,t)}{\partial t}=D\frac{\partial^2}{\partial x^2}\Big( \big(1-r(x)\big)\omega(x)p(x,t) \Big),
        \label{eq:DTRWMarkovian_continuummaster}
\end{equation}
with the diffusion limit
$
        D=\lim_{\substack{\Delta x\to 0\\\Delta t\to 0}}\frac{\Delta x^2}{2\Delta t}.
$
In the Markovian DTRW, the probability of jumping on a given time step at site $i$ is $\omega(i)$ and the probability that the jump is not a self-jump at site $i$ is $1-r(i)$.
The product of these terms is the probability that the particle jumps away from site $i$ in one time step.
This leads to the identification of a space-dependent characteristic waiting time
\begin{equation}
        \tau(x)=\frac{\Delta t}{\big(1-r(x)\big)\omega(x)},
\end{equation}
so that \linkref[equation]{eq:DTRWMarkovian_continuummaster} can be rewritten in the form of \linkref[equation]{eq:CTRWMarkov_continuummaster} as
\begin{equation}
        \frac{\partial p(x,t)}{\partial t}=\frac{\partial^2}{\partial x^2}\Big(D(x)p(x,t)\Big),
        \label{eq:DTRWMarkovian_continuummaster2}
\end{equation}
with a space-dependent diffusion coefficient arising from a global diffusion limit,
\begin{equation}
        D(x)= \lim_{\substack{\Delta x\to 0\\\Delta t\to 0}} \big(1-r(i\Delta x)\big)\omega(i\Delta x) \frac{\Delta x^2}{2\Delta t}.
\end{equation}
Note that in the above $1-r(x)$ and $\omega(x)$ play similar roles, such that $\big(1-r(x)\big)\omega(x)$ is the effective probability that particles jump away from the position $x$ in one time step. This changes if the waiting step mass function is no longer Markovian, and $1-r(x)$ plays a more important role in obtaining a global diffusion limit.

\subsection{Space-dependent non-Markovian waiting step mass function}
\label{sec:DTRWnonmarkov}

For the non-Markovian DTRW case, we consider a Sibuya waiting time PMF,
\begin{equation}
        \psi(i,n) = \begin{cases}
                0, & \text{if }n\leq 0,\\
                \displaystyle \frac{\alpha(i)}{n}\prod_{k=1}^{n-1}\left(1-\frac{\alpha(i)}{k}\right), & \text{if } n>0.
        \end{cases}
        \label{eq:DTRWNonmarkovwaitingtime}
\end{equation}
\linkref[Equation]{eq:DTRWMarkovwaitingtime} gives the survival probability \linkref{eq:DTRWsurvivalprobability}
\begin{equation}
        \Psi(i,n) = \prod_{k=1}^{n}\left(1-\frac{\alpha(i)}{k}\right),
        \label{eq:DTRWNonmarkovsurvival}
\end{equation}
and kernel \linkref{eq:DTRWKernelZtransformed},
\begin{equation}
        K(i,n)=\left(-1\right)^n\binom{1-\alpha(i)}{n}+\delta_{n,1}-\delta_{n,0}.
        \label{eq:DTRWNonmarkovkernel}
\end{equation}
In this non-Markovian case, the expected waiting time diverges.
If $\alpha$ is constant, then the expected number of jumps in $n$ time steps scales as $n^{\alpha}$ \cite{nichols2018subdiffusive} such that one could define a characteristic waiting time $\tau = n^{\alpha}\Delta t$ globally.
If $\alpha(x)$ varies in space, a characteristic waiting time can only be defined locally.
In the absence of self-jumping, i.e. $r(i)=0, \forall i$, the master equation for the non-Markovian DTRW with a space varying order $\alpha(x)$ could be written as \linkref[equation]{eq:vofde} with
\begin{equation}\label{kappaxup}
        \kappa_{\alpha(x)}=\lim_{\substack{\Delta t\to 0\\\Delta x\to 0}}
        \frac{\Delta x^2}{2\Delta t^{\alpha(x)}}.
\end{equation}
However, as discussed previously, this limit is unphysical as it cannot converge to a finite non-zero value for all $x$ unless $\alpha$ is a constant.

So now, we consider the inclusion of self-jumping.
Using the kernel \linkref{eq:DTRWNonmarkovkernel} in the general DTRW master \linkref[equation]{eq:DTRWmasterRestNN}, we obtain
\begin{equation}
        \begin{split}
                P(i,n)
                -P(i,n-1) = &
                \left(\frac{1-r(i-1)}{2}\right) P(i-1,n-1)
                + \left(\frac{1-r(i+1)}{2}\right)P(i+1,n-1)
                \\
                &
                - (1-r(i))P(i,n-1)
                \\
                &
                +\sum_{m=0}^{n-1} (-1)^{n-m}\left[
                \frac{1-r(i-1)}{2} \binom{1-\alpha(i-1)}{n-m} P(i-1,m)
                \right.\\
                &
                + \frac{1-r(i+1)}{2} \binom{1-\alpha(i+1)}{n-m} P(i+1,m)
                \\
                &
                \left.
                -\left(1-r(i)\right)\binom{1-\alpha(i)}{n-m} P(i,m) \right].
        \end{split}
        \label{DTRWNonmarkovmaster}
\end{equation}
From \linkref[equation]{DTRWNonmarkovmaster}, we can derive the variable-order fractional diffusion \linkref[equation]{eq:vofde} through a general method using the $\mathcal{Z}^*$-transform.
The main tools and some technical aspects are included in the text below but the full details are provided in appendix~\ref{sec:Ztransform}.

We introduce the $\mathcal{Z}^*$-transform in time as
\begin{equation}
        \mathcal{Z}_n^*[Y(i,n)|s\Delta t]=\sum_{n=0}^\infty e^{-sn\Delta t}Y(i,n),
\end{equation}
with the convolution property
\begin{equation}
        \begin{split}
                \mathcal{Z}_n^*
                &
                [X(i,n)|s\Delta t] \, \mathcal{Z}_n^*[Y(i,n)|s\Delta t]
                =
                \mathcal{Z}_n^*\left[\sum_{m=0}^n X(i,m)Y(i,n-m)|s\Delta t\right]
        \end{split}
\end{equation}
and the time shift property,
\begin{equation}
        \mathcal{Z}_n^*[Y(i,n\pm 1)|s\Delta t]=e^{\pm s\Delta t}\mathcal{Z}_n^*[Y(i,n)|s\Delta t].
\end{equation}
After taking the $\mathcal{Z}^*$-transform in time and space of \linkref[equation]{DTRWNonmarkovmaster} and expanding the exponential terms for small $\Delta t$ and $\Delta x$, we obtain
\begin{equation}
        \begin{split}
                s &\Delta t \hat{\mathcal{Z}}_i^* \left[\mathcal{Z}_n^*\left[P(i,n)|s\Delta t\right] |q\Delta x\right] \approx
                \frac{q^2\Delta x^2}{2}\hat{\mathcal{Z}}_i^*\left[ \left( 1-r(i) \right) \left(s\Delta t\right)^{1-\alpha(i)} \mathcal{Z}_n^*\left[P(i,n)|s\Delta t\right] \bigg | q\Delta x\right].
        \end{split}
        \label{eq:ZstarTransformDTRW}
\end{equation}
Further details can be found in appendix~\ref{sec:Ztransform}.

After some calculations, we can evaluate the integrals to obtain the variable-order fractional diffusion \linkref[equation]{eq:vofde}
\begin{equation}
        \frac{\partial p(x,t)}{\partial t} = \frac{\partial^2}{\partial x^2} \kappa_{\alpha(x)} {}_{0}{\mathcal{D}}_t^{1-\alpha(x)} p(x,t),
        \label{eq:vofde_dtrw}
\end{equation}
where
\begin{equation}
        \kappa_{\alpha(x)} = \lim_{\substack{\Delta t\to 0\\\Delta x\to 0}} \frac{\Delta x^2}{2\Delta t^{\alpha(x)}} \big(1-r(x)\big).
        \label{eq:diffcoeff_nonmarkovDTRW}
\end{equation}
The steps moving from \linkref[equation]{eq:ZstarTransformDTRW} to \linkref[equation]{eq:vofde_dtrw} are detailed in appendix~\ref{sec:Ztransform}.

Through the inclusion of self-jumping in \linkref[equation]{DTRWNonmarkovmaster}, we can obtain a well-defined diffusion coefficient if we set the space-dependent single step rest probability, $r(x)$ in \linkref[equation]{eq:diffcoeff_nonmarkovDTRW} as
\begin{equation}
        1-r(x) = \frac{\Delta t^{\alpha(x)}}{t_0^{\alpha(x)} \tau},
        \label{eq:rfactor}
\end{equation}
where $t_0$ has dimensions of time and $\tau$ is dimensionless. In this case, the fractional diffusion coefficient is
\begin{equation}
        \kappa_{\alpha(x)} = \lim_{\substack{\tau\to 0\\\Delta x\to 0}} \frac{\Delta x^2}{2t_0^{\alpha(x)}\tau} = D t_0^{-\alpha(x)}.
        \label{eq:DTRWkappaCorrect}
\end{equation}
Since $r(x)$ is a probability, $r(x)\leq 1$ requires $t_0^{\alpha(x)}\tau \geq \Delta t^{\alpha^*}$, where $\alpha^* = \min\left(\alpha(x)\right)$.
This provides a relation between the continuum time limit scaling factor $\Delta t$ and the diffusion limit scaling factor $\tau$, which is important for simulations of the DTRW.
In particular, without loss of generality, we find that we can set $\tau=\Delta t^{\alpha^*}$.
This particular choice corresponds to taking the equality above and setting $t_0=1$. We can set $t_0=1$ without loss of generality because results for the variable-order time-fractional diffusion equation, with $t_0\ne 1$, can be obtained from time rescaled solutions of the variable-order diffusion equation, with $t_0=1$. This is shown in appendix~\ref{sec:t0_choice}.

So in deriving a global diffusion limit \linkref{eq:DTRWkappaCorrect} from the DTRW with self-jumping, we have set the probability of a self-jump as the ratio between a position dependent absolute time scale $\Delta t^{\alpha(x)}$ and a position dependent effective characteristic time scale $t_0^{\alpha(x)}\tau$.
This effective characteristic time scale is composed of a local part $t_0^{\alpha(x)}$ and a global part $\tau$ and allows a global diffusion limit to be taken.
In the CTRW, the effective characteristic time scale was introduced into the Mittag--Leffler waiting time density \linkref{eq:MLwaitingtimes_rescaledspacedep} directly as an initial assumption.

We remark that within an unbiased nearest-neighbour DTRW, the self-jump is the minimal modification that allows a global diffusion limit.
It arises naturally from ensuring conservation of the probability mass from transitions: what remains after left and right neighbour jumps must be reassigned to the original site.
It is possible that other more complicated schemes could allow a global diffusion limit.
However, different schemes would result in different governing PDEs.
For example, introducing a bias between neighbouring sites or including jumps outside nearest neighbours could accommodate global diffusion limits but would modify the spatial operator.
The discreteness of waiting times in the DTRW naturally leads to modification of the transition PMF and the need to reproduce the Laplacian operator is what forces us to introduce self-jumps as opposed to any other strategy to obtain a global diffusion limit.

It is also worth noting that while our derivation has considered a Sibuya distribution for the discrete waiting time PMF, other heavy-tailed discrete probability distributions could be explored. Examples include the discrete Pareto distribution and the Yule distribution. These three distributions have the same heavy-tailed behaviour \cite{kozubowski2018generalized}
\[
\mathbb{P}(N=n)=O\left(\frac{1}{n^{\alpha+1}}\right) \quad \mbox{as}\quad n\to \infty.
\]

The steps above using the $\mathcal{Z}^*$-transform are general and can be applied for any DTRW given a suitable kernel.
This general method gives a systematic way to move from DTRW master equations to the governing partial differential equation of the RW density in the continuum limit.
In appendix~\ref{sec:HD}, we show that the same techniques can be applied to derive the variable-order time-fractional equation in higher dimensions.

\section{Numerical methods and Monte Carlo simulations}
\label{sec:numericalMethods}

Now, we show how the non-Markovian DTRW master equation with self-jumping, \linkref[equation]{eq:DTRWmasterRestNN} with the kernel defined by \linkref[equation]{eq:DTRWNonmarkovkernel}, can be adapted to provide a numerical method for solving the variable-order fractional diffusion \linkref[equation]{eq:vofde}.
We begin by identifying
\begin{equation}
        p(x,t)\approx P(i,n),
\end{equation}
where $i=\lfloor\frac{x}{\Delta x}\rfloor$ and $n=\lfloor\frac{t}{\Delta t}\rfloor$.
Then, we can specify the lattice spacing and time increment by using \linkref[equations]{eq:DTRWkappaCorrect} and \linkref{eq:diffcoeff_nonmarkovDTRW}
\begin{equation}
        \begin{split}
                \Delta x^2
                                = 2\kappa_{\alpha(x)}t_0^{\alpha(x)}\tau
                                \quad \text{and} \quad
                \Delta t
                                = \left[ \frac{\Delta x^2}{2\kappa_{\alpha(x)}}(1-r(x)) \right]^{\frac{1}{\alpha(x)}}.
        \end{split}
\end{equation}
Since $1-r(x)$ is a probability, \linkref[equation]{eq:rfactor} implies that $t_0^{\alpha(x)}\tau \geq \Delta t^{\alpha_*}$, where $\alpha_* = \min\left(\alpha(x)\right)$.
Then, we can choose $\Delta t = t_0^{\alpha(x)/\alpha_*}\tau^{1/\alpha_*}$.
For convenience, we can choose $t_0=1$ and consequently choose $\kappa_{\alpha(x)} = 1$.
This gives
\begin{equation}
        \Delta x^2 / 2 = \tau \quad \text{and} \quad \Delta t = \tau^{\frac{1}{\alpha_*}} = \left(\frac{\Delta x^2}{2}\right)^{\frac{1}{\alpha_*}}.
\end{equation}
All that remains is to fix $\Delta x$ or $\Delta t$ depending on the desired outcome of the numerical method.
To obtain a solution up to time $T,$ we need to evolve the DTRW master equation, \linkref[equation]{DTRWNonmarkovmaster} for $N=\lfloor\frac{T}{\Delta t}\rfloor$ steps.
Although the preceding sections have only considered the infinite lattice, the finite domain case can be easily derived in the same fashion and boundary conditions can be implemented in the standard way (see for example \cite{angstmann2015discrete}).

It is also possible to formulate a Monte Carlo simulation for the CTRW through random walking particles that wait at a lattice site $i$ for a time $T$ drawn from a Mittag--Leffler waiting time density \linkref{eq:MLwaitingtimes} \cite{fedotov2012subdiffusive,fedotov2019asymptotic} before jumping to a neighbouring site.
To obtain a global diffusion limit, our analysis above leads to a modified Monte Carlo scheme compared to existing literature.
We first define a lattice spacing $\Delta x = L/k$ where $k+1$ counts the number of equally spaced points, including end points, on the interval $x\in[0,L]$.
The space-dependent fractional exponent is discretized to $\alpha(i) = \alpha(i\Delta x)$ for $i = 0, \cdots, k$ and the minimum value of the exponent is set to $\alpha_*$.
Then, we define the parameter $\tau = \Delta x^2/2$ and subsequently the discrete rest probability $r(i) = r(i\Delta x),$ where $r(x) = 1-\tau^{(\alpha(x)-\alpha_*)/\alpha_*}$.
The discrete rest probability, $r(i),$ is the probability that the waiting time of a random walker expires in state $i$ and the random walker remains at $i$ for another Mittag--Leffler distributed waiting time.

For each stochastic realization, the particle initially starts according to an initial condition at $t=0$ and follows the steps:
\begin{enumerate}
        \item[(i)] The particle waits a time $T$ drawn from a Mittag--Leffer density that can be generated by (see \S9.2 in \cite{kozubowski1999univariate})
        \begin{equation}
                T=-\tau \log(U)\left(\frac{\sin(\alpha(x)\pi)}{\tan(\alpha(x)\pi V)}-\cos(\alpha(x)\pi)\right)^{\frac{1}{\alpha(x)}},
        \end{equation}
        where $U\text{ and }V$ are random numbers distributed uniformly on $(0,1)$.
        \item[(ii)] The particle then remains at the current location, $x$, with probability $r(x)$ and moves to one of the nearest neighbours, $x\pm \Delta x$, with probability $(1-r(x))/2$. Reflective barriers are imposed at $x=0$ and $x=L$, consistent with \cite{feller1968};
        if a particle reaches an end point $x=0$, or $x=L$, then on the next jump it has equal probability to remain at that end point or jump to $x=\Delta x$ or $x=L-\Delta x$, respectively.
        \item[(iii)] The time is incremented to $t \leftarrow t + T$ and spatial position is recorded.
        \item[(iv)] Steps (i)--(iii) are repeated until a desired time $t\geq t_{\mathrm{end}}$ is reached.
\end{enumerate}
This provides a numerical approximation for the space--time behaviour of $p(x,t)$ where
\begin{equation}
        \frac{\partial p(x,t)}{\partial t}
        =\frac{\partial^2}{\partial x^2}\left(\frac{\Delta x^2}{2\tau t_0^{\alpha(x)}}\, _0\mathcal{D}_t^{1-\alpha(x)}\right)p(x,t),
\end{equation}
and $t_0=1$ has dimensions of time.
The Monte Carlo approximation becomes better for smaller $\Delta x$ with $\tau = \frac{\Delta x^2}{2}$ as here $\lim_{\substack{\tau\to 0\\\Delta x\to 0}} \Delta x^2 / 2\tau_0$ does exist and is finite.
It is also possible to implement a Monte Carlo simulation for the DTRW directly by generating Sibuya distributed random variables, see \cite{leonenko2025sibuya} for methods.

\begin{figure}[htbp]
\centering
\includegraphics[width=0.8\linewidth]{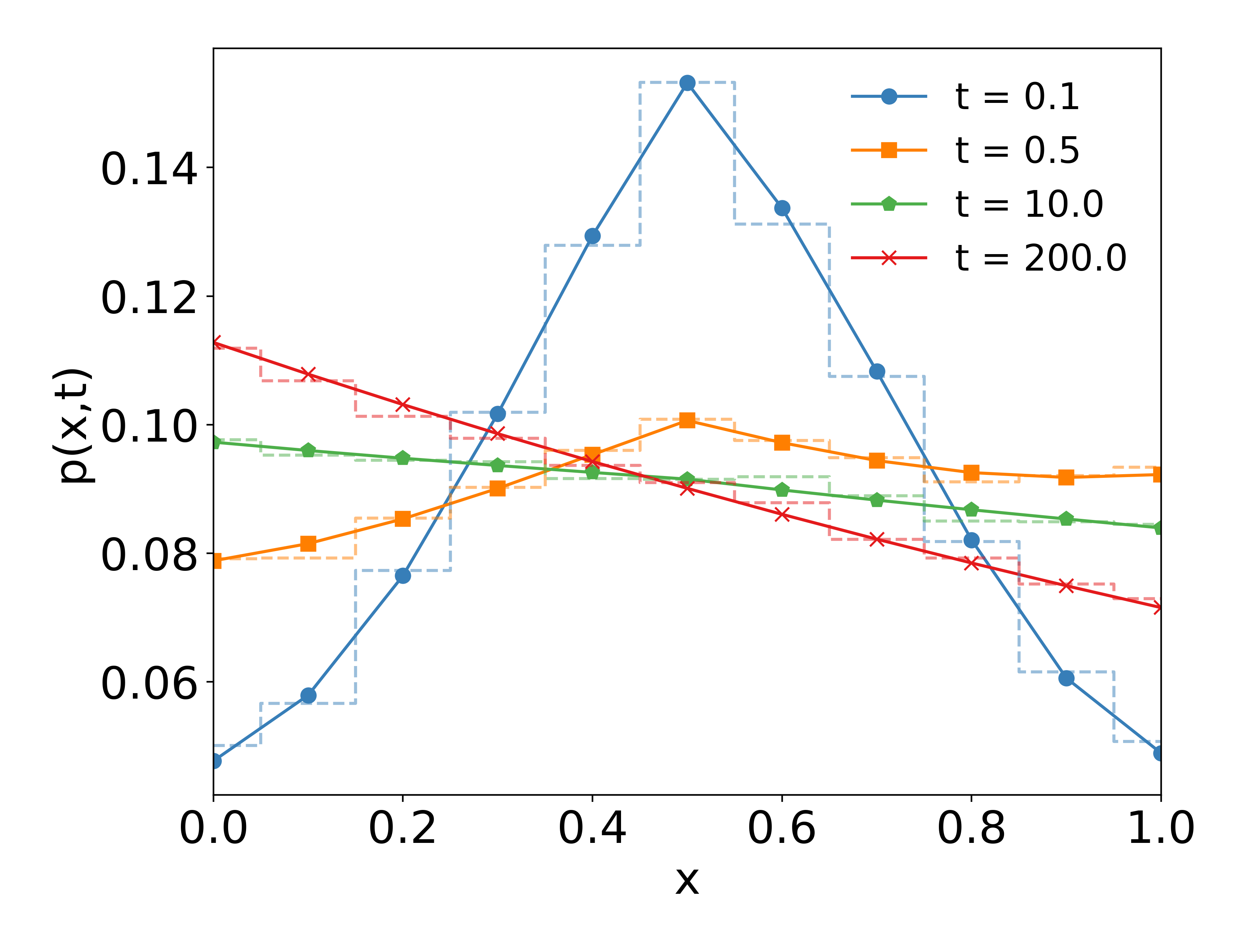}
\caption{Plot of the PDFs, $p(x,t)$, calculated via numerical integration of the DTRW (solid lines) and Monte Carlo simulation (dashed lines). Different colours show different times. The space-dependent exponent was $\alpha(x) = 0.8+0.1x$ with $x\in[0,1]$ and $11$ discrete spatial lattice points such that $\Delta x=0.1$. The initial condition was $p(x,t) = \delta(x-0.5)$.}
\label{fig:dtrw_mc_1}
\end{figure}

Figure~\ref{fig:dtrw_mc_1} shows results from Monte Carlo simulations, using the above scheme, with $N=11$ points and $\Delta x=0.1$ and $\tau_0=\frac{{\Delta x}^2}{2}$ on a lattice $x\in [0,1]$.
Note that $\alpha(x)$ and $p(x,t)$ share the same lattice spacing of $\Delta x$ in figure~\ref{fig:dtrw_mc_1} with $\alpha(x)$ being a continuous function. In general, this need not be the case because if $\alpha(x)$ is piecewise constant, then the spatial discretization can be finer than the spatial scale at which $\alpha(x)$ changes.
In these simulations, each particle initially starts at the midpoint of the lattice with the time set at $t=0$ with $n=0$.
The variable-order exponent is $\alpha(x)=0.8+0.1x$, and the probability to remain at the same site after a Mittag--Leffler random waiting time $T$ is $r(x)=1-\tau_0^{(\alpha(x)-\alpha^*)/\alpha^*}$, where $\alpha^*=0.8$.
Monte Carlo simulations have been carried out for 100\,000 particles and the relative numbers of particles at lattice sites are shown in histogram form for times $t=0.1, 0.5, 10$ and $200$.
The results of the Monte Carlo simulations (dashed lines) are in excellent agreement with the DTRW simulations (markers and solid lines).
We see the approach towards a symmetric distribution before the anomalous aggregation towards $x \in \text{argmin}(\alpha(x))$ dominates \cite{fedotov2019asymptotic}.

\section{Discussion}

Advances in transport measurements at microscopic levels, particularly in biological media, have revealed complex diffusion processes that cannot be accurately modelled with standard RWs and standard diffusion equations.
There has been much theoretical progress over the past few decades exploiting properties of CTRWs and fractional calculus.
This is now well understood in systems that are spatially homogeneous on mesoscopic scales where time-fractional derivatives have become part of a modelling paradigm for systems in which diffusion is anomalously slow owing to traps and obstacles.
Diffusion equations, Fokker--Planck equations and reaction diffusion equations with time-fractional derivatives are now common place in mathematical modelling.
One of the lessons that has been learned, and is still being learnt in this field, is that it may not be possible to obtain a well-defined diffusion model by naively replacing ordinary time derivatives with fractional derivatives.
The various models need to be physically justified, usually through an underlying RW.
There are also derivations of the anomalous diffusion equation arising from Noether's second theorem \cite{baggioli2021anomalous}.

Here, we have considered the possibility of using time-fractional derivatives, with spatially varying order, in models for diffusion in systems that are spatially inhomogeneous on mesoscopic scales.
We have shown that a space-dependent variable-order time-fractional diffusion equation can be derived from a CTRW with an appropriately scaled Mittag--Leffler waiting time density that has a space-dependent parameter $\alpha(x)$.
The additional parameters are a dimensionless scale factor $\tau$ and a parameter $t_0$ with dimensions of time that appears as $t_0^{\alpha(x)}$ in the denominator of the diffusion coefficient.
The parameter $t_0$ provides consistent physical dimensions through a spatially varying derivative in time.
If the time-fractional derivative is constant in space, $\alpha(x)=\alpha$, then $t_0^\alpha$ can be taken as a global characteristic waiting time that can be taken to zero, in a diffusion limit, to balance the lattice spacing $\Delta x$ being taken to zero for the spatial continuum limit.
If $\alpha(x)$ is not constant in space then a global diffusion limit cannot be obtained in this way.
Rather than using a waiting time as a scale parameter, it is the dimensionless parameter $\tau$ that is taken to zero in a diffusion limit.

To better understand this diffusion limit, we derived the variable-order time-fractional diffusion equation from a DTRW.
The fractional diffusion equation with a constant order time-fractional derivative had been derived from a DTRW previously \cite{angstmann2015discrete} with waiting times distributed by a Sibuya PMF.
One of the advantages in dealing with a DTRW is that the discrete time automatically provides an additional parameter, the discrete time interval $\Delta t$ which can be taken to zero to realize continuous time, as the lattice spacing $\Delta x$ is taken to zero to realize continuous space.
Then, we introduced a probability for the DTRW to jump out of a state and return, the self-jump, which depended on $\Delta t$ and spatial position.
With this approach, we were able to obtain a well-defined space- and time-continuum limit with a globally defined diffusion coefficient.
We recruited the additional dimensionless scale factor $\tau$ and the additional local characteristic waiting time $t_0^{\alpha(x)}$ in the definition for the probability of a self-jump.
We showed that solutions can be obtained from numerical methods based on the master equation for the DTRW with a Sibuya waiting time PMF.
Further, we introduced a modified Monte Carlo scheme whose results were in excellent agreement with the numerical method.

Our results provide a physically consistent justification for the consideration of the space-dependent variable-order time-fractional diffusion equation as a mathematical model for diffusive transport in spatially inhomogeneous systems with power law waiting time traps, characterized by an exponent $\alpha(x)$. One of the signature features of this model is symmetry breaking in which the spreading of an initially spatially symmetric plume transitions towards an accumulation at spatial locations, where $\alpha(x)$ is a minimum. The experimental systems, identified in the introduction \cite{pieprzyk2016spatially,bruna2015diffusion,stickler2011continuous,marbach2022mass,zheng2024hopping,bressloff2013stochastic,han2020deciphering,fedotov2021variable,waigh2023heterogeneous,ambegaokar1971hopping,egami2012spacetime,henri2023control,roupas2021gravitational}, are examples of systems that could be investigated for this sort of behaviour.

Two natural generalizations lie outside the present scope of the paper and are interesting and important questions.
Replacing the nearest-neighbour jumps with heavy-tailed L\'evy jump lengths \cite{metzler2000random} or with Sibuya compounding \cite{angstmann2025compounded} would introduce Riesz or spectral space-fractional derivatives, respectively.
This would yield an underlying RW model that gives rise to a space-dependent variable-order time-fractional diffusion equation with an additional space-fractional derivative.
More complicated scaling questions would appear if the fractional exponents associated with the time and space-fractional derivatives were spatially varying.
In a similar vein, the Riemann--Liouville time-fractional derivative appears as a consequence of the generalized diffusion limit from the underlying RW.
While the Caputo fractional derivative \cite{caputo1967linear} can be used instead for constant fractional exponents, this interchangeability disappears when the fractional exponent is spatially varying.
It would be insightful if an underlying RW could be constructed such that Caputo time-fractional derivatives of variable order appear from the generalized diffusion limit.

\section*{Data accessibility}
This article has no additional data.

\section*{Declaration of AI use}
AI-assisted technologies were used to improve readability and language of the work.

\section*{Authors' contributions}
C.N.A.: conceptualization, investigation, methodology, writing---review and editing; D.S.H.: conceptualization, formal analysis, investigation, methodology, software, validation, visualization, writing---original draft, writing---review and editing; B.I.H.: conceptualization, formal analysis, investigation, methodology, visualization, writing---original draft, writing---review and editing; B.Z.H.: investigation, software, validation, writing---review and editing; Z.X.: investigation, methodology, software, validation, writing---review and editing.

All authors gave final approval for publication and agreed to be held accountable for the work performed therein.

\section*{Conflict of interest declaration}
We declare we have no competing interests.

\section*{Funding}
This research was funded by Australian Research Council grant number DP200100345 and UNSW Science Faculty Research Grants.

\section*{Acknowledgements}
We thank S. Burney, W. Li and A. McGann for useful discussions and the anonymous referees for their constructive feedback.

\appendix

\section{Continuous time random walk generalized master equation}
\label{appA}

In this appendix, for completeness, we provide details leading from the CTRW generalized master equation to the evolution equations in \linkref[equations]{eq:CTRWmaster}, \linkref{eq:integral_escape_rate} and \linkref{eq:master_nonmarkov}.
For generality, we consider a CTRW with nearest-neighbour unbiased jumps on a lattice with spacing $\Delta x$, and with a space-dependent waiting time density $\psi(x,t)$ and a space-dependent survival probablity
\begin{equation}\label{Psi}
        \Psi(x,t)=1-\int_0^t \psi(x,t')\, dt'.
\end{equation}
The generalized master equation for this can be written as \cite{angstmann2015generalized}
\begin{equation}\label{Agme}
        \begin{split}
                \frac{\partial P(x,t)}{\partial t} ={}& \frac{1}{2}\int_0^t K(x-\Delta x,t-t')P(x-\Delta x,t') dt'
                +\frac{1}{2}\int_0^t K(x+\Delta x,t-t')P(x+\Delta x,t')\, dt'\\
                &-\int_0^t K(x,t-t')P(x,t')\, dt',
        \end{split}
\end{equation}
where the kernel, $K(x,t)$ is defined in Laplace space, as the ratio between the waiting time density and survival function:
\begin{equation}
        \mathcal{L}_t \left\{ K(x,t) \right\}(s) = \frac{       \mathcal{L}_t \left\{ \psi(x,t) \right\} (s)}{        \mathcal{L}_t \left\{ \Psi(x,t) \right\} (s)}.
\end{equation}
Here, $\mathcal{L}_t\left\{ \cdot \right\}(s)$ denotes the Laplace transform with respect to time $t$ and $s$ is the transform variable.
Note that using \linkref[equation]{Psi}, we can write
\[
\mathcal{L}_t \left\{ \Psi(x,t) \right\} (s)=\frac{1}{s}-\frac{\mathcal{L}_t \left\{ \psi(x,t) \right\} (s)}{s}
\]
and then
\begin{equation}\label{LK}
        \mathcal{L}_t \left\{ K(x,t) \right\}(s) = \frac{       s\mathcal{L}_t \left\{ \psi(x,t) \right\} (s)}{1-      \mathcal{L}_t \left\{ \psi(x,t) \right\} (s)}.
\end{equation}

We now consider a Mittag--Leffler waiting time density of the form
\begin{equation}
        \psi(x,t)= \frac{t^{\alpha(x)-1}}{\tau^{\alpha(x)}}
        E_{\alpha(x),\alpha(x)}
        \left(
        -\left(
        \frac{t}{\tau}
        \right)^{\alpha(x)}
        \right),
\end{equation}
with $0<\alpha(x)<1$.
The Laplace transform in this case, which can be found using eqn (2.2.21) from \cite{mathai2008special}, is given by
\begin{equation}
        \mathcal{L}_t \left\{ \psi(x,t) \right\} (s)=\frac{1}{1+(s\tau)^{\alpha(x)}}
\end{equation}
and then using \linkref[equation]{LK}, we have
\[
\mathcal{L}_t \left\{ K(x,t) \right\}(s)=\frac{s^{1-\alpha(x)}}{\tau^{\alpha(x)}}.
\]

It now follows from the convolution result for Laplace transforms that
\[
\mathcal{L}_t \left\{\int_0^t K(x,t-t') y(x,t')\, dt'\right\}(s)=\frac{s^{1-\alpha(x)}}{\tau^{\alpha(x)}}
\mathcal{L}_t\left\{ y(x,t)\right\}(s).
\]
But we also have that
\[
\mathcal{L}_t \left\{ \, _0\mathcal{D}_t^{1-\alpha(x)} y(x,t)\right\}(s)=s^{1-\alpha(x)}
\mathcal{L}_t\left\{ y(x,t)\right\}(s)
\]
provided that the fractional integral $_0\mathcal{D}_t^{-\alpha(x)}y(x,t)$ vanishes at $t=0$.

We can now write
\begin{equation}
        \int_0^t K(x,t-t') y(x,t')\, dt'=\frac{1}{\tau^{\alpha(x)}}\, _0\mathcal{D}_t^{1-\alpha(x)} y(x,t)
\end{equation}
so that we can replace the integrals in \linkref[equation]{Agme} with fractional derivatives to arrive at
\begin{equation}
        \begin{split}
                \frac{\partial P(x,t)}{\partial t} ={}& \frac{1}{2\tau^{\alpha(x-\Delta x)}}\, _0\mathcal{D}_t^{1-\alpha(x-\Delta x)} P(x-\Delta x,t)
                +\frac{1}{2\tau^{\alpha(x+\Delta x)}}\, _0\mathcal{D}_t^{1-\alpha(x+\Delta x)} P(x+\Delta x,t)\\
                &-\frac{1}{\tau^{\alpha(x)}}\, _0\mathcal{D}_t^{1-\alpha(x)} P(x,t),
        \end{split}
\end{equation}
which recovers the result in \linkref[equation]{eq:master_nonmarkov}.
If $\alpha(x)=\alpha$ is a constant then this simplifies to
\begin{equation}
        \begin{split}
                \frac{\partial P(x,t)}{\partial t} = \frac{1}{2\tau^{\alpha}}\, _0\mathcal{D}_t^{1-\alpha} P(x-\Delta x,t)
                +\frac{1}{2\tau^{\alpha}}\, _0\mathcal{D}_t^{1-\alpha} P(x+\Delta x,t)
                -\frac{1}{\tau^{\alpha}}\, _0\mathcal{D}_t^{1-\alpha} P(x,t),
        \end{split}
\end{equation}
which recovers the results in \linkref[equations]{eq:CTRWmaster} and \linkref{eq:integral_escape_rate}.

\section{\texorpdfstring{$\mathcal{Z}^*$}{Z*}-transform and derivation of governing equation}
\label{sec:Ztransform}

We introduce the $\mathcal{Z}^*$-transform in time as
\begin{equation}
        \mathcal{Z}_n^*[Y(i,n)|s\Delta t]=\sum_{n=0}^\infty e^{-sn\Delta t}Y(i,n),
\end{equation}
with the convolution property
\begin{equation}
        \begin{split}
                \mathcal{Z}_n^*[X(i,n)|s\Delta t] \, \mathcal{Z}_n^*[Y(i,n)|s\Delta t]&=
                \sum_{n=0}^\infty X(i,n)e^{-sn\Delta t}
                \sum_{n=0}^\infty Y(i,n)e^{-sn\Delta t}\\
                &=
                \sum_{n=0}^\infty\sum_{m=0}^n X(i,m)Y(i,n-m)e^{-sn\Delta t}\\
                &=\mathcal{Z}_n^*\left[\sum_{m=0}^n X(i,m)Y(i,n-m)|s\Delta t\right]
        \end{split}
\end{equation}
and the time shift property,
\begin{equation}
        \mathcal{Z}_n^*[Y(i,n\pm 1)|s\Delta t]=e^{\pm s\Delta t}\mathcal{Z}_n^*[Y(i,n)|s\Delta t].
\end{equation}
After taking the $\mathcal{Z}^*$-transform in time across \linkref[equation]{DTRWNonmarkovmaster}, we obtain,
\begin{align}
       & \left(1-e^{-s\Delta t}\right)\mathcal{Z}_n^*\left[P(i,n) |s\Delta t\right]=
        -\left(\frac{1-r(i-1)}{2}\right)
        \left(1-e^{-s\Delta t}\right)
        \mathcal{Z}_n^*\left[P(i-1,n)|s\Delta t\right]\nonumber\\
        & \quad -\left(\frac{1-r(i+1)}{2}\right)
        \left(1-e^{-s\Delta t}\right)
        \mathcal{Z}_n^*\left[P(i+1,n)|s\Delta t\right]\nonumber\\
        & \quad  +\left(1-r(i)\right)
        \left(1-e^{-s\Delta t}\right)
        \mathcal{Z}_n^*\left[P(i,n)|s\Delta t\right]\nonumber\\
        & \quad +\left(\frac{1-r(i-1)}{2}\right)
        \mathcal{Z}_n^*\left[P(i-1,n)|s\Delta t\right]
        \mathcal{Z}_n^*\left[(-1)^n \binom{1-\alpha(i-1)}{n}\bigg |s\Delta t\right]\nonumber\\
        & \quad  +\left(\frac{1-r(i+1)}{2}\right)
        \mathcal{Z}_n^*\left[P(i+1,n)|s\Delta t\right]
        \mathcal{Z}_n^*\left[(-1)^n\binom{1-\alpha(i+1)}{n}\bigg | s\Delta t\right]\nonumber\\
        & \quad -\left(1-r(i)\right)
        \mathcal{Z}_n^*\left[P(i,n)|s\Delta t\right]
        \mathcal{Z}_n^*\left[(-1)^n \binom{1-\alpha(i)}{n}\bigg |s\Delta t\right].
        \label{SSSmaster}
\end{align}
Next, we introduce the $\hat{\mathcal{Z}}^*$-transform in space as
\begin{equation}
        \hat{\mathcal{Z}}_i^*[Y(i,n)|q\Delta x]=\sum_{i=-\infty}^\infty e^{-qi\Delta x}Y(i,n),
\end{equation}
with the space shift property
\begin{equation}
        \hat{\mathcal{Z}}_i^*[Y(i\pm 1,n)|q,\Delta x]=
        e^{\pm q\Delta x}\hat{\mathcal{Z}}_i^*[Y(i,n)|q\Delta x].
\end{equation}
After taking the $\hat{\mathcal{Z}}^*$-transform in space of \linkref[equation]{SSSmaster}, we obtain
\begin{align}
        &\left(1-e^{-s\Delta t}\right)\hat{\mathcal{Z}}_i^*\left[\mathcal{Z}_n^*\left[P(i,n)|s\Delta t\right] |q\Delta x\right] \nonumber\\
        & \quad = -\left( \frac{1}{2}e^{-q\Delta x}+\frac{1}{2}e^{q\Delta x}-1 \right) \hat{\mathcal{Z}}_i^*\left[
        \left( 1-r(i) \right)
        \left(1-e^{-s\Delta t}\right)
        \mathcal{Z}_n^*\left[P(i,n)|s\Delta t\right]
        \bigg | q\Delta x\right]  \nonumber\\
        & \quad \,+ \left( \frac{1}{2}e^{-q\Delta x}+\frac{1}{2}e^{q\Delta x}-1 \right) \hat{\mathcal{Z}}_i^*\left[
        \left( 1-r(i) \right)
        \mathcal{Z}_n^*\left[P(i,n)|s\Delta t\right]
        \mathcal{Z}_n^*\left[(-1)^n \binom{1-\alpha(i)}{n} \bigg | s\Delta t\right]
        \Bigg | q\Delta x\right].
        \label{eq:DTRWSpaceTimeZstarTransformed}
\end{align}
We now use the result
\begin{equation}
        \mathcal{Z}_n^*\left[ (-1)^n \binom{1-\alpha(i)}{n} \bigg | s\Delta t \right] = \left(1-e^{-s\Delta t}\right)^{1-\alpha(i)}
\end{equation}
to write \linkref[equation]{eq:DTRWSpaceTimeZstarTransformed} as
\begin{equation}
        \begin{split}
                &\left(1-e^{-s\Delta t}\right)\hat{\mathcal{Z}}_i^*\left[\mathcal{Z}_n^*\left[P(i,n)|s\Delta t\right] |q\Delta x\right] \\
                & \quad = -\left( \frac{1}{2}e^{-q\Delta x}+\frac{1}{2}e^{q\Delta x}-1 \right) \hat{\mathcal{Z}}_i^*\left[
                \left( 1-r(i) \right)
                \left(1-e^{-s\Delta t}\right)
                \mathcal{Z}_n^*\left[P(i,n)|s\Delta t\right]
                \bigg | q\Delta x\right] \\
                & \quad  + \left( \frac{1}{2}e^{-q\Delta x}+\frac{1}{2}e^{q\Delta x}-1 \right) \hat{\mathcal{Z}}_i^*\left[
                \left( 1-r(i) \right)
                \left(1-e^{-s\Delta t}\right)^{1-\alpha(i)}
                \mathcal{Z}_n^*\left[P(i,n)|s\Delta t\right]
                \bigg | q\Delta x\right]. \\
        \end{split}
        \label{eq:DTRWSpaceTimeZstarTransformedBinomSimplified}
\end{equation}
We now expand the exponential terms for small $\Delta t$ and $\Delta x$ to obtain
\begin{equation}
        s\Delta t \hat{\mathcal{Z}}_i^*\left[\mathcal{Z}_n^*\left[P(i,n)|s\Delta t\right] |q\Delta x\right] \approx \frac{q^2\Delta x^2}{2}\hat{\mathcal{Z}}_i^*\left[
        \left( 1-r(i) \right)
        \left(s\Delta t\right)^{1-\alpha(i)}
        \mathcal{Z}_n^*\left[P(i,n)|s\Delta t\right]
        \bigg | q\Delta x\right].
\end{equation}
Note that the term multiplied by $\left(1-e^{-s\Delta t}\right)$ in \linkref{eq:DTRWSpaceTimeZstarTransformed} and \linkref{eq:DTRWSpaceTimeZstarTransformedBinomSimplified} vanishes in the limit as $\Delta t\rightarrow 0$ as they are dominated by the term multiplied by $\left(1-e^{-s\Delta t}\right)^{1-\alpha(i)}$ since $0<\alpha(i)<1$.
Then,
\begin{align}
       & \lim_{\substack{\Delta t\to 0\\\Delta x\to 0}} s\sum_{i=-\infty}^{\infty} e^{-iq\Delta x}
        \sum_{n=0}^\infty e^{-ns\Delta t}p_\Delta(i\Delta x,n\Delta t)\Delta t  \nonumber\\
        &\quad \approx \lim_{\substack{\Delta t\to 0\\\Delta x\to 0}} \sum_{i=-\infty}^\infty \frac{\Delta x^2}{2\Delta t^{\alpha(i\Delta x)}}q^2e^{-iq\Delta x}(1-r(i\Delta x)) s^{1-\alpha(i\Delta x)} \sum_{n=0}^\infty e^{-ns\Delta t}p_\Delta(i\Delta x,n\Delta t)\Delta t.
        \label{eq:DTRWapproxLimit}
\end{align}
If we identify $x=i\Delta x$ and $t = n\Delta t$, then \linkref[equation]{eq:DTRWapproxLimit} becomes
\begin{align}
        & s \int_{-\infty}^{\infty} e^{-qx'}
        \left(\int_0^\infty e^{-st'}p(x',t')\, dt'\right) dx'
         \nonumber\\
        &\quad  \approx \int_{-\infty}^{\infty}
        \left(\lim_{\substack{\Delta t\to 0\\\Delta x\to 0}}\frac{\Delta x^2}{2\Delta t^{\alpha(x')}} \right)
        (1-r(x'))
        q^2e^{-qx'}s^{1-\alpha(x')}
        \left( \int_0^\infty e^{-st'} p(x',t')\, dt'\right) dx'.
\label{equ13}
\end{align}
We now take the inverse Laplace transform in space and time to arrive at
\begin{align}
       & \frac{\partial }{\partial t}
        \int_{-\infty}^{\infty}\delta(x-x')
        \left(\int_0^\infty \delta(t-t')p(x',t')\, dt'\right) dx'
        \nonumber\\
        &\quad  \approx \frac{\partial^2}{\partial x^2}
        \int_{-\infty}^{\infty}
        \left(\lim_{\substack{\Delta t\to 0\\\Delta x\to 0}}\frac{\Delta x^2}{2\Delta t^{\alpha(x')}} \right)
        (1-r(x')) \delta(x-x')
        _0\mathcal{D}_t^{1-\alpha(x')}
        \left( \int_0^\infty \delta(t-t') p(x',t') dt'\right) dx'.
        \label{eq:inverseLaplace_master}
\end{align}
In obtaining \linkref[equation]{equ13} from \linkref[equation]{eq:DTRWapproxLimit}, we have used an inverse unilateral Laplace transform in time and an inverse bilateral Laplace transform in space. The introduction of the Dirac delta function in \linkref[equation]{eq:inverseLaplace_master} can be justified in two ways, either through the replacements $f(t)=\int_0^\infty \delta(t-t') f(t')\, dt'$ and $g(x)=\int_{-\infty}^\infty \delta(x-x') g(x')\, dx'$ after performing the inverse transforms, or by writing $f(t)=1\times f(t)$ and $g(x)=1\times g(x)$ before taking the inverse transform and then using the convolution theorem in taking the inverse transform with the further identification that the unilateral, or bilateral, transform of a Dirac delta function is unity.
Finally, we can evaluate the integrals to obtain the variable-order fractional diffusion equation
\begin{equation}
        \frac{\partial p(x,t)}{\partial t} = \frac{\partial^2}{\partial x^2} \kappa_{\alpha(x)} {}_{0}{\mathcal{D}}_t^{1-\alpha(x)} p(x,t),
\end{equation}
where
\begin{equation}
        \kappa_{\alpha(x)} = \lim_{\substack{\Delta t\to 0\\\Delta x\to 0}} \big(1-r(x)\big)\frac{\Delta x^2}{2\Delta t^{\alpha(x)}}.
        \label{eq:diffcoeff_nonmarkovDTRW_appendix}
\end{equation}

\section{Variable-order fractional diffusion equation in higher dimensions}
\label{sec:HD}

The derivation of the variable-order time-fractional diffusion equation from a DTRW above can be applied to higher dimensions by considering a lattice $V\in\mathbb{Z}^{\mathfrak{D}}$ with $z_i$, $L_{1,i}$, $L_{2,i} \in \mathbb{Z}$ for $i = 1,2,\cdots,{\mathfrak{D}}$.

Then, we can rewrite the DTRW master \linkref[equation]{eq:DTRW_generalmaster} as
\begin{equation}
        \begin{split}
                P(\vec{z},n)-P(\vec{z},n-1)
                =
                \sum_{\vec{y}\in V}\sum_{m=0}^{n-1}w(\vec{z}|\vec{y})K(\vec{y},n-m)P(\vec{y},m)
                - \sum_{m=0}^{n-1}K(\vec{z},n-m)P(\vec{z},m),
        \end{split}
\end{equation}
where $\vec{z}= (z_1,z_2,\cdots,z_{\mathfrak{D}})$.

The transition PMF \linkref{eq:DTRWrighttransitionPMF} in ${\mathfrak{D}}$ dimensions becomes
\begin{equation}
        \begin{split}
                w(\vec{z},\vec{y}) =
                \frac{1-r(\vec{z})}{2{\mathfrak{D}}} \left[ \sum_{i=1}^{\mathfrak{D}} \delta(\vec{z}-\vec{y} + \vec{e}_i) + \sum_{i=1}^{\mathfrak{D}} \delta(\vec{z}-\vec{y} - \vec{e}_i)\right]
                + r(\vec{z})\delta(\vec{z}-\vec{y}),
        \end{split}
\end{equation}
where $\vec{e}_i$ is a unit vector in the $i$th component in $V$. Using this transition PMF, the DTRW master \linkref[equation]{DTRWNonmarkovmaster} in higher dimensions can be written as
\begin{align}
 &  P(\vec{z},n)-P(\vec{z},n-1) \nonumber \\
 &\quad  = \frac{1}{2{\mathfrak{D}}} \sum_{i=1}^{\mathfrak{D}}\left[ \left(1-r(\vec{z}+\vec{e}_i)\right)P(\vec{z}+\vec{e}_i,n-1) + \left(1-r(\vec{z}-\vec{e}_i)\right)P(\vec{z}-\vec{e}_i,n-1) \right]  \nonumber\\
  & \quad  - (1-r(\vec{z}))P(\vec{z},n-1)\nonumber \\
 &\quad  +\frac{1}{2{\mathfrak{D}}}\sum_{m=0}^{n-1} (-1)^{n-m} \sum_{i=1}^{\mathfrak{D}} (1-r(\vec{z}+\vec{e}_i)) \binom{1-\alpha(\vec{z}+\vec{e}_i)}{n-m} P(\vec{z}+\vec{e}_i,m)    \nonumber \\
  &\quad  +  \frac{1}{2{\mathfrak{D}}}\sum_{m=0}^{n-1} (-1)^{n-m} \sum_{i=1}^{\mathfrak{D}}  (1-r(\vec{z}-\vec{e}_i)) \binom{1-\alpha(\vec{z}-\vec{e}_i)}{n-m} P(\vec{z}-\vec{e}_i,m) \nonumber \\
 &\quad  - \frac{1}{2{\mathfrak{D}}}\sum_{m=0}^{n-1} (-1)^{n-m} (1-r(\vec{z})) \binom{1-\alpha(\vec{z})}{n-m} P(\vec{z},m).
 \label{DTRWNonmarkovmasterND}
\end{align}

Then, following the same steps as in \S\ref{sec:DTRWnonmarkov} and appendix~\ref{sec:Ztransform}, and using the spatial $\mathcal{Z}^*$-transform repeatedly for the ${\mathfrak{D}}$ dimensions, we arrive at the ${\mathfrak{D}}$-dimensional variable-order time-fractional diffusion equation
\begin{equation}
        \partial_t p(\vec{z},t) = \sum_{i=1}^{\mathfrak{D}}\partial^2_{z_i}\left( \kappa_{\alpha(\vec{z})}^{z_i} {}_{0}{\mathcal{D}}_t^{1-\alpha(\vec{z})} p(\vec{z},t) \right),
        \label{eq:DdimensionalVOFDE}
\end{equation}
where the diffusion coefficients are
\begin{equation}
        \kappa_{\alpha(\vec{z})}^{z_i} = \lim_{\substack{\Delta t\to 0\\\Delta z_i\to 0, \forall i}}\frac{\Delta z_i^2}{2\mathfrak{D}\Delta t^{\alpha(\vec{z})}}(1-r(\vec{z})).
\end{equation}
If we define the probability to leave site $\vec{z}$ as
\begin{equation}
        1-r(\vec{z}) = \frac{\Delta t^{\alpha(\vec{z})}}{t_0^{\alpha(\vec{z})} \tau},
\end{equation}
then a global diffusion limit can be taken as previously seen for the one-dimensional case. For a uniform lattice spacing in all orthogonal directions, $\Delta z_i = a$ for some constant $a>0$, the ${\mathfrak{D}}$-dimensional variable-order time-fractional diffusion \linkref[equation]{eq:DdimensionalVOFDE} has an isotropic diffusion coefficient and simplifies to
\begin{equation}
        \partial_t p(\vec{z},t) = \Delta \kappa_{\alpha(\vec{z})} {}_{0}{\mathcal{D}}_t^{1-\alpha(\vec{z})} p(\vec{z},t).
        \label{eq:DdimensionalVOFDE_isotropic}
\end{equation}

\section{\texorpdfstring{Choice of arbitrary $t_0$}{Choice of arbitrary t0}}
\label{sec:t0_choice}

In this appendix, we show that if $p_0(x,t)$ denotes the solution of
\begin{equation}\label{canon}
        \frac{\partial p(x,t)}{\partial t}=D\frac{\partial^2}{\partial x^2}\, _0\mathcal{D}_t^{1-\alpha(x)}p(x,t),
\end{equation}
with initial condition $p_0(x,0)=\delta(x)$, and $p(x,t;t_0)$ denotes the solution of
\begin{equation}\label{spaceD}
        \frac{\partial p(x,t)}{\partial t}=\frac{\partial^2}{\partial x^2} \left(\kappa_\alpha(x)\, _0\mathcal{D}_t^{1-\alpha(x)} p(x,t)\right),
\end{equation}
with the same initial condition, then $p(x,\tilde{t};t_0)=p_0(x,t),$ where $\tilde{t} = t t_0$.

To see this, note that if $p(x,t;t_0)$ satisfies \linkref[equation]{spaceD} for all times, then $p(x,\tilde{t};t_0)$ satisfies the equation
\begin{equation}
        \frac{\partial p(x,\tilde{t};t_0)}{\partial t}=\frac{\partial^2}{\partial x^2}D t_0^{-\alpha(x)}
        \frac{1}{\Gamma(\alpha(x))}\frac{d}{dt}\int_0^{\tilde{t}}
        \frac{p(x,t';t_0)dt'}{(\tilde{t}-t')^{1-\alpha(x)}}
\end{equation}
and then
\begin{equation}
        \frac{\partial p(x,\tilde{t};t_0)}{\partial t}=\frac{\partial^2}{\partial x^2}D
        \frac{1}{\Gamma(\alpha(x))}\frac{1}{t_0}\frac{d}{dt}\int_0^{\tilde{t}}
        \frac{p(x,t';t_0)dt'}{(t-\frac{t'}{t_0})^{1-\alpha(x)}}.
\end{equation}
Now make a change of variables $t''=t'/ t_0$, then
\begin{equation}
        \frac{\partial p(x,\tilde{t};t_0)}{\partial t}=\frac{\partial^2}{\partial x^2}D
        \frac{1}{\Gamma(\alpha(x))}\frac{d}{dt}\int_0^{t}
        \frac{p(x,t'' t_0;t_0)dt''}{(t-t'')^{1-\alpha(x)}}.
\end{equation}
Let $\tilde p(x,t;t_0)=p(x,\tilde{t};t_0)$ and use $t'$ as the dummy integration variable then
\begin{equation}
        \frac{\partial\tilde p(x,t;t_0)}{\partial t}=\frac{\partial^2}{\partial x^2}D
        \frac{1}{\Gamma(\alpha(x))}\frac{d}{dt}\int_0^{t}
        \frac{\tilde p(x,t';t_0)dt'}{(t-t')^{1-\alpha(x)}},
\end{equation}
which is the evolution equation for $p_0(x,t)$ that does not depend on $t_0$.
Thus, $p(x,\tilde{t};t_0)$ and $p_0(x,t)$ satisfy the same evolution equation and they are equal if they satisfy the same initial condition.
The result $p(x,\tilde{t};t_0)=p_0(x,t)$ was demonstrated numerically for a particular value of $t_0$ using Monte Carlo simulations in \cite{straka2018variable}.
It also agrees with the asymptotic result in \cite{fedotov2019asymptotic} and the result was established in \cite{roth2020inhomogeneous} using a different approach.
This is a fundamentally important result for applications of the variable-order time-fractional diffusion equation. We can derive results for the canonical variable-order diffusion equation, with $t_0=1$, and results for the more general variable-order time-fractional diffusion equation, with $t_0\ne 1$, follow from a simple rescaling of time.

\section{Further numerical results}
\label{sec:furtherNumerics}

\begin{figure}[htbp]
\centering
\includegraphics[width=0.8\linewidth]{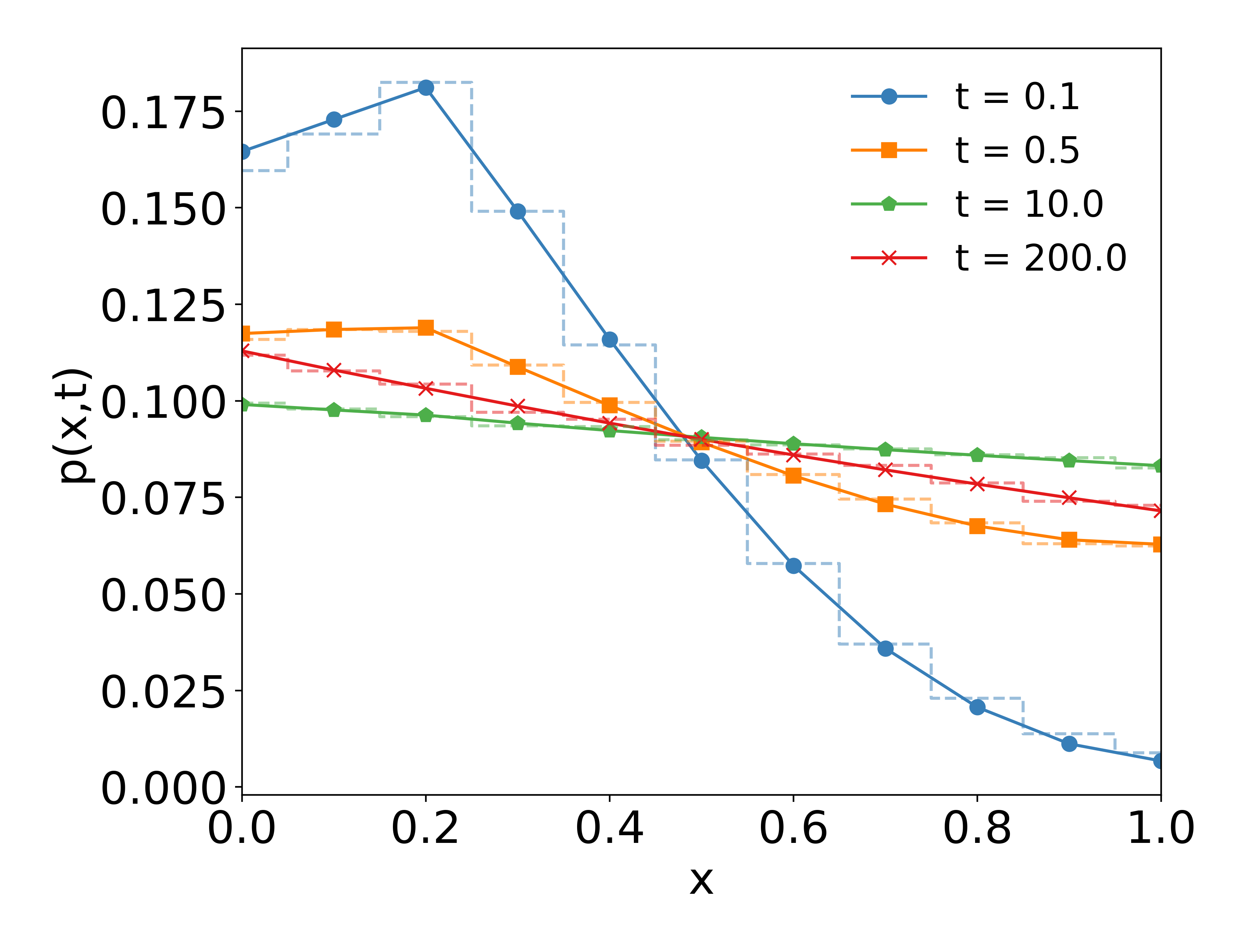}
\caption{Plot of the PDFs, $p(x,t)$, with the same parameters as figure~\ref{fig:dtrw_mc_1} except the initial condition was $p(x,t)=\delta(x-0.2)$.}
\label{fig:dtrw_mc_2}
\end{figure}

\begin{figure}[htbp]
\centering
\includegraphics[width=0.8\linewidth]{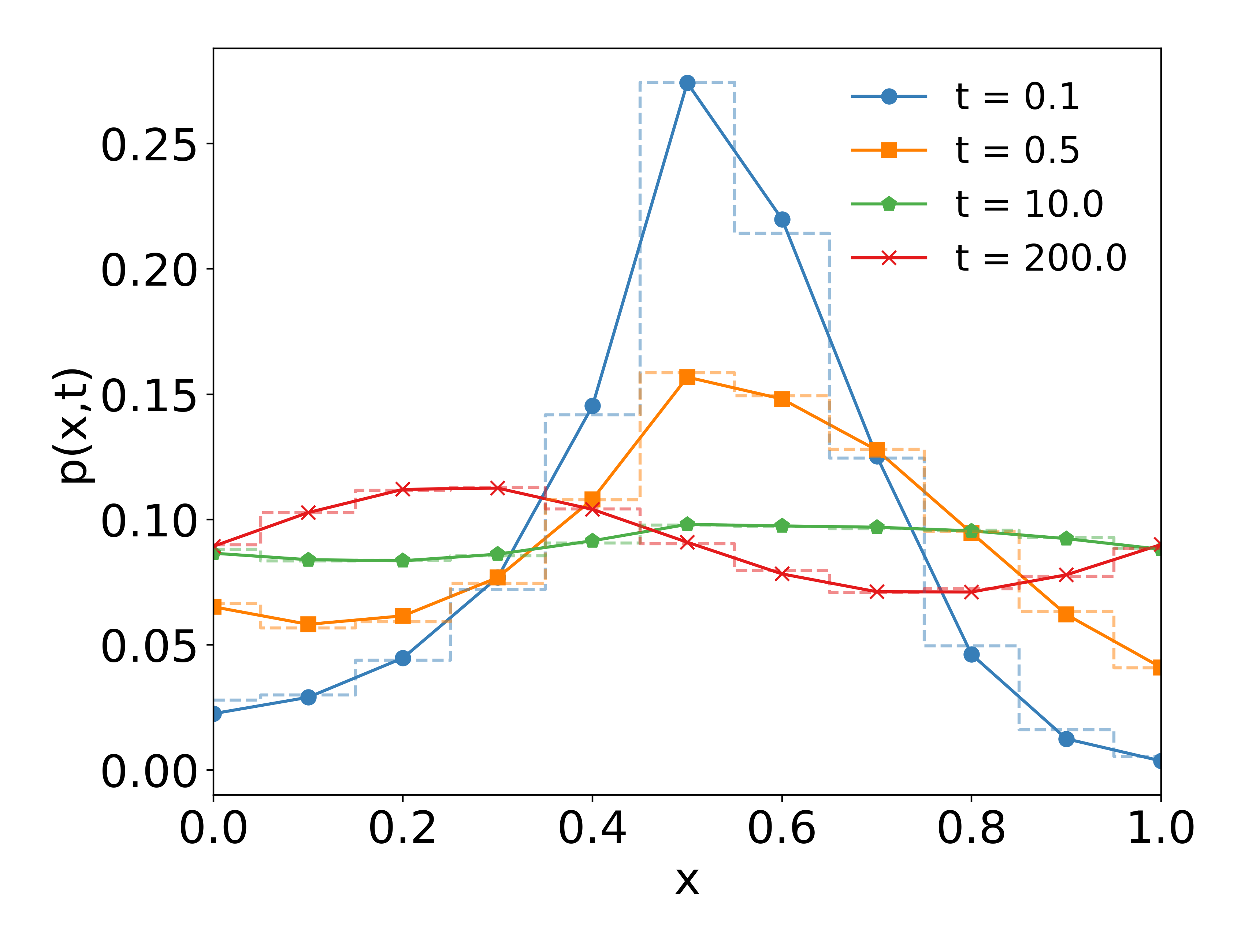}
\caption{Plot of the PDFs, $p(x,t)$, with the same parameters as figure~\ref{fig:dtrw_mc_1} except the fractional exponent is $\alpha(x) = 0.7-0.1\sin(2\pi x)$.}
\label{fig:dtrw_mc_3}
\end{figure}

\begin{figure}[htbp]
\centering
\includegraphics[width=0.8\linewidth]{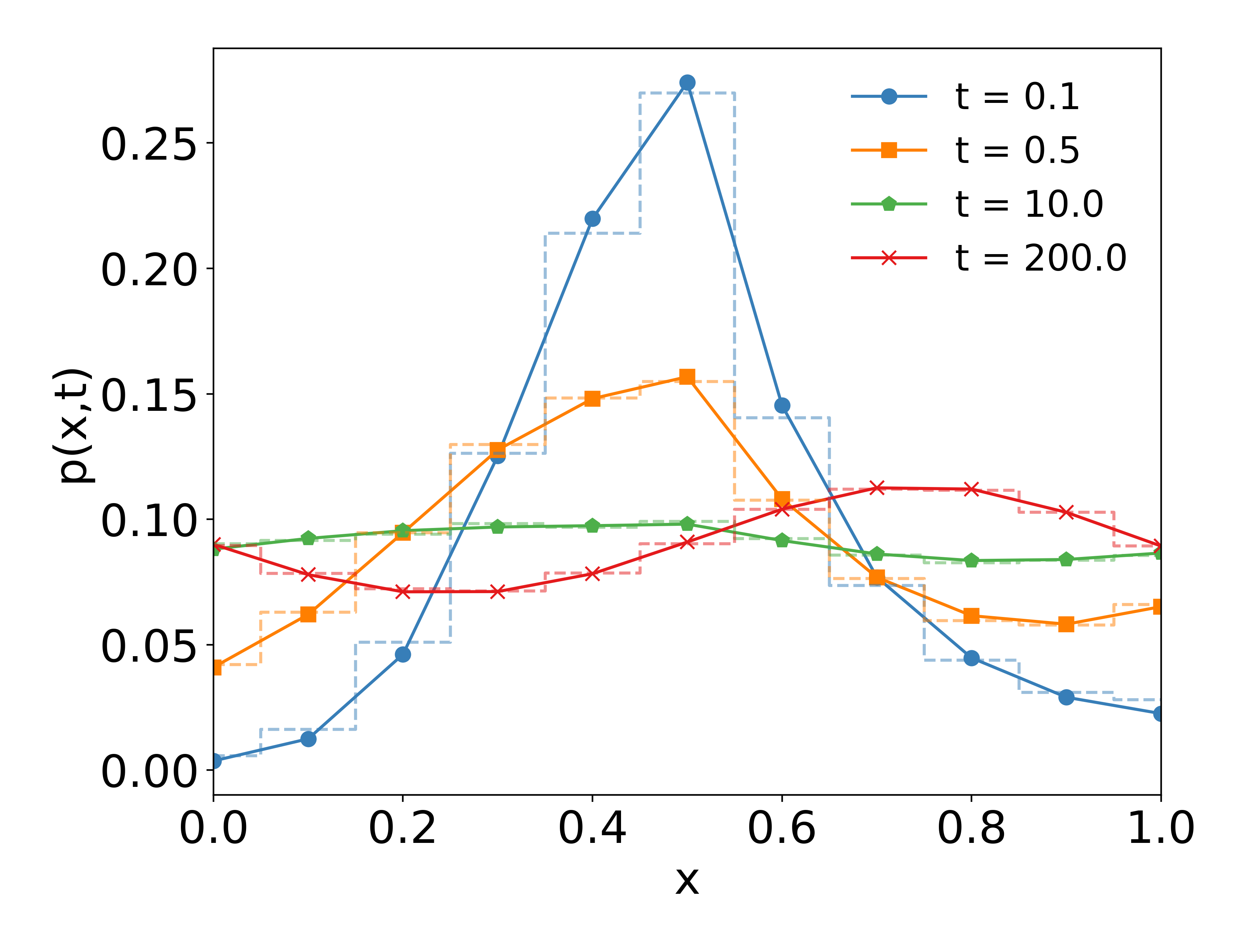}
\caption{Plot of the PDFs, $p(x,t)$, with the same parameters as figure~\ref{fig:dtrw_mc_1} except the fractional exponent is $\alpha(x) = 0.7+0.1\sin(2\pi x)$.}
\label{fig:dtrw_mc_4}
\end{figure}

Figures~\ref{fig:dtrw_mc_2}, \ref{fig:dtrw_mc_3} and \ref{fig:dtrw_mc_4} show the results of DTRW numerical approximations and Monte Carlo simulations of the underlying RW in continuous time for different initial conditions and different fractional exponent, $\alpha(x)$. The agreement of numerical approximations with Monte Carlo simulations demonstrates further evidence of the validity and consistency of both methods.

\end{document}